\documentclass{aa} 

\usepackage{cmap}
\usepackage{amsmath,amssymb,amstext}
\usepackage{babel}
\usepackage{times}
\usepackage[Bjarne]{fncychap} 
\usepackage{geometry}
\usepackage{graphicx} \graphicspath{{img/}}
\usepackage{xspace}
\usepackage{hyphenat}
\usepackage[ruled,vlined,resetcount,linesnumbered]{algorithm2e}
\usepackage{listings}
\usepackage{natbib}
\usepackage{xcolor}
\usepackage{appendix}
\usepackage{mathrsfs}
\usepackage{verbatim}
\usepackage{multirow}
\usepackage{booktabs}

\newcommand{\Pp}{P}

\newcommand{\Dl}{\Delta\ell}
\newcommand{\al}{\alpha^*}
\newcommand{\tz}{(\alpha^*, z)}
\newcommand{\tp}{(\alpha^*, \Pp)}

\def\rc{r}
\def\tr{\tau_{\rc}^\lambda}

\usepackage{hyperref}
\hypersetup{colorlinks=true,citecolor=blue,linkcolor=red!80!black,urlcolor=blue!80!black}
\usepackage{hypcap}
\newcommand{\cref}[1]{Fig.~\ref{#1}}

\newcommand{\taurex}{TauREx\xspace}
\newcommand{\pytmo}{Pytmosph3R\xspace}

\usepackage[colorinlistoftodos]{todonotes}

\begin{document}
	
\title{Toward a multidimensional analysis of transmission spectroscopy.
Part III: Modelling 2D effects in retrievals with \taurex\thanks{The \taurex~2D plugin is available at \url{https://forge.oasu.u-bordeaux.fr/falco/taurex_2d}}}
\titlerunning{Modelling 2D effects in retrievals with \taurex}

\author{Tiziano Zingales\inst{1, 2}
\and
Aur\'{e}lien Falco\inst{1}
\and
William Pluriel\inst{1, 3}
\and
J\'{e}r\'{e}my Leconte\inst{1}
}

\institute{Laboratoire d'Astrophysique de Bordeaux, Univ. Bordeaux, CNRS, B18N, all\'{e}e Geoffroy Saint-Hilaire, 33615 Pessac, France 
\and 
Universit\`a di Padova, Dipartimento di Astronomia, vicolo dell’Osservatorio 3, 35122 Padova, Italy 
\and
Observatoire astronomique de l’Universit\'e de Gen\`eve, chemin Pegasi 51, 1290 Versoix, Switzerland}

\date{\today}

\abstract{New-generation spectrographs dedicated to the study of exoplanetary atmospheres require a high accuracy in the atmospheric models to better interpret the input spectra. Thanks to space missions like James Webb Space Telescope (JWST), ARIEL and Twinkle, the observed spectra will indeed cover a large wavelength range from visible to mid-infrared with an higher precision compared to the old-generation instrumentation, revealing complex features coming from different regions of the atmosphere. For hot and ultra hot Jupiters (HJs and UHJs), the main source of complexity in the spectra comes from thermal and chemical differences between the day and the night sides. In this context, one-dimensional plane parallel retrieval models of atmospheres may not be suitable to extract the complexity of such spectra. 
In addition, Bayesian frameworks are computationally intensive and prevent us from using complete three-dimensional self-consistent models to retrieve exoplanetary atmospheres, and they constrain us to use simplified models to converge to a set of atmospheric parameters. 
We thus propose the \taurex~2D retrieval code, which uses two-dimensional atmospheric models as a good compromise between computational cost and model accuracy to better infer exoplanetary atmospheric characteristics for the hottest planets.
\taurex~2D uses a 2D parametrization across the limb which computes the transmission spectrum from an exoplanetary atmosphere assuming azitmuthal symmetry. It also includes a thermal dissociation model of various species. 
We demonstrate that, given an input observation, \taurex~2D mitigates the biases between the retrieved atmospheric parameters and the real atmospheric parameters. We also show that having a prior knowledge on the link between local temperature and composition is instrumental in inferring the temperature structure of the atmosphere.  Finally, we apply such a model on a synthetic spectrum computed from a GCM simulation of WASP-121b and show how parameter biases can be removed when using two-dimensional forward models across the limb.

}

\maketitle

\section{Introduction}
\label{introduction}

Exoplanets have diverse and complex atmospheres, which often require very sophisticated methods to be correctly interpreted. In particular, hot Jupiters (HJs) and ultra hot Jupiters (UHJs) appear to have a strong dichotomy between the inflated hot day side and the cold and shrunken night side \citep{showmann2002,showman2008,menou2009,wordsworth2011,heng2011,charnay2015,kataria2016, drummond2016, tan2019,pluriel2020a,Pluriel2022}. For such planets, it is decisive to use more-than-one dimensional models to, at least, reproduce realistically the different chemistry and physical properties.

Transit spectroscopy allows the observer to infer atmospheric features by looking at a limited volume around the terminator line \citep{brown2001, kreidberg2018}. The transmitted light carries the information of both the day and the night side of the planet of a small region around the terminator line \citet{fortney2010, caldas2019effects, Wardenier2022}. For cold planets, a one-dimensional plane parallel geometry can be enough to interpret low resolution data since the atmosphere is more homogeneous \citep{MacDonald2020}. Nevertheless, in the case of HJs and UHJs, the use of a higher-dimensional model becomes crucial to mitigate the biases between the retrieved model and the real three dimensional observed atmosphere. The impact of 3D atmospheric structures cannot be neglected and will have an impact on the spectra thus on the retrieval analysis because of chemistry heterogeneities \citep{Changeat_2019, Baeyens2021}, or due to morning-evening asymmetry \citep{line2016,parmentierr2016,MacDonald2020,espinoza2021}.

Atmospheric retrievals encompass a series of algorithms which try to associate an input exoplanetary spectrum to an atmospheric model \citep{rodgers2000,irwin2008,madhusudhan2009,line2013,waldmann2015a,waldmann2015b,gandhi2018,al2019taurex,molliere2019,zhang2019,min2020}. Bayesian frameworks are very useful to find the best atmospheric models which will reproduce the input spectra. They are often the main analysis tool to infer atmospheric properties of exoplanets. The use of such tools poses two main challenges:

\begin{itemize}
	\item Computational cost: Bayesian analysis is typically computationally-intensive, especially when one tries to fit several model parameters (e.g. higher than 10). In order to reach convergence in a few hours timespan, it is necessary to use parallelized-software on a multi-core environment.
	\item Model accuracy: Real atmospheres are complex three-dimensional structures, whose physics can be computationally-intensive to reproduce correctly. Moreover, such complex 3D atmospheres often lead to degeneracies when using too broad assumptions in the models.
\end{itemize}

Inverse modeling requires the fast generation of a high number of forward models (e.g. $10^3-10^7$). Each forward model thus needs to be computed in a very short amount of time to allow the Bayesian Framework to converge in a reasonable amount of time. 
For this reason the retrieval community is obliged to accept very strong assumptions on the atmospheric model to find a good compromise between computational time and the atmospheric model's precision, given the accuracy of an input spectrum.\\
The Hubble/WFC3 camera represents the state-of-the-art technology for the study of exoplanetary atmospheres, sensitive in the 1.1$\mu$m-1.7$\mu$m wavelength range, with an accuracy down to $\sim$50~ppm. With the information given by the Hubble/WFC3 camera, a 1D plane parallel atmosphere can be enough to claim detection of several molecular species present in the exoplanetary spectra, i.e. H$_2$O, NH$_3$, CH$_4$, TiO and VO \citep{deming2010,macdonald2017,molaverdikhani2019,changeat2021}. It is however challenging to get to elementary abundances i.e., O/H, C/O etc., only with the information from the WFC3 wavelength range.
Today, there are several studies which demonstrate the presence of water using Hubble data with a precision up to 0.3 dex \citep{wakeford2018,tsiaras2018,welbanks2019,barstow2020}.

Future space missions, like ARIEL \citep{ariel2016,ariel2018}, Twinkle \citep{twinkle2016} and JWST \citep{jwst2006}, will spend part or most of their observation time on exoplanets. 
They will have a wider wavelength range as well as a higher spectral resolution and spectral accuracy \citep{beichman2014}. Planets like HJs and UHJs, which will be observed by these future space missions cannot be well interpreted by merely assuming a 1D plane parallel atmosphere as a forward model \citep{pluriel2020a}. More complex atmospheric models will be necessary to explain some spectral features correctly, and a compromise between computational cost and model accuracy will still be required.
There has been several attempts to address these challenges, taking into considerations, for example, phase curves retrievals \citep{feng2016,changeat2020phase,feng2020,taylor2020,lacy2020,MacDonald2020,MacDonald2021trident,espinoza2021, mikalevans2022}.\\
All of these factors lead us to reconsider the use of one dimensional models and to reconstruct the forward model used in the \taurex atmospheric retrieval tool \citep{waldmann2015a,waldmann2015b,Al-Refaie2021} considering an additional geometric dimension. More-than-1D models and retrievals have recently been discussed by \cite{lacy2020,MacDonald2021trident} and very recently by \cite{Nixon2022aura3D}. In particular, the approach of \citet{lacy2020} relies on a similar 2D atmospheric structure, based on \citet{caldas2019effects}. However, their chemical approach differs from our parametrization. They consider chemical equilibrium everywhere and retrieve the bulk metallicity. In \taurex~2D, we can use two modes (see Sec. \ref{parameters}); one with more freedom: the  abundances  in  both  hemispheres are retrieved separately. The second mode assumes a chemical equilibrium, taking into account thermal dissociation, and we retrieve the deep abundance of each species. 
Another difference between our approach and that of \citet{lacy2020} is our use of a deep temperature parameter set for pressures higher than a threshold. These 2 parameters, based on hot Jupiter GCM simulations, correspond to the level where the radiative time scale becomes higher than the dynamics time scale e.g. to an homogeneous temperature layer.

In the following sections, we propose a two-dimensional atmospheric model as a good compromise between model accuracy and computational requirement for a Bayesian analysis. The model, discussed in its forward usage in \citet{falco2020taurex2D}, is tested here within a Bayesian framework. We show the advantages and the weaknesses of using a more-than-one dimensional model through a benchmark of three atmospheric configurations with increasing complexity. Finally, we investigate the atmosphere of an ultra hot Jupiter, following the results of \citet{pluriel2020a}, in presence of molecular dissociation, to quantify the intrinsic model biases introduced within a Bayesian framework. One of the main characteristics of \taurex~2D is the introduction of compositional variation between day and night side produced by thermal dissociation of atmospheric molecules.

\section{Two-dimensional Model}
\label{model-presentation}

The forward model used in \taurex~2D is formally described in \citet{falco2020taurex2D}. We give here a short presentation of the model and refer to the above paper for more information.

The model relies on the assumption that the atmosphere is symmetrical around the star-observer axis. The computation of the transmission spectrum therefore reduces to a two-dimensional problem.

\begin{figure}[ht]
	\centering	
	\includegraphics[width=.4\textwidth]{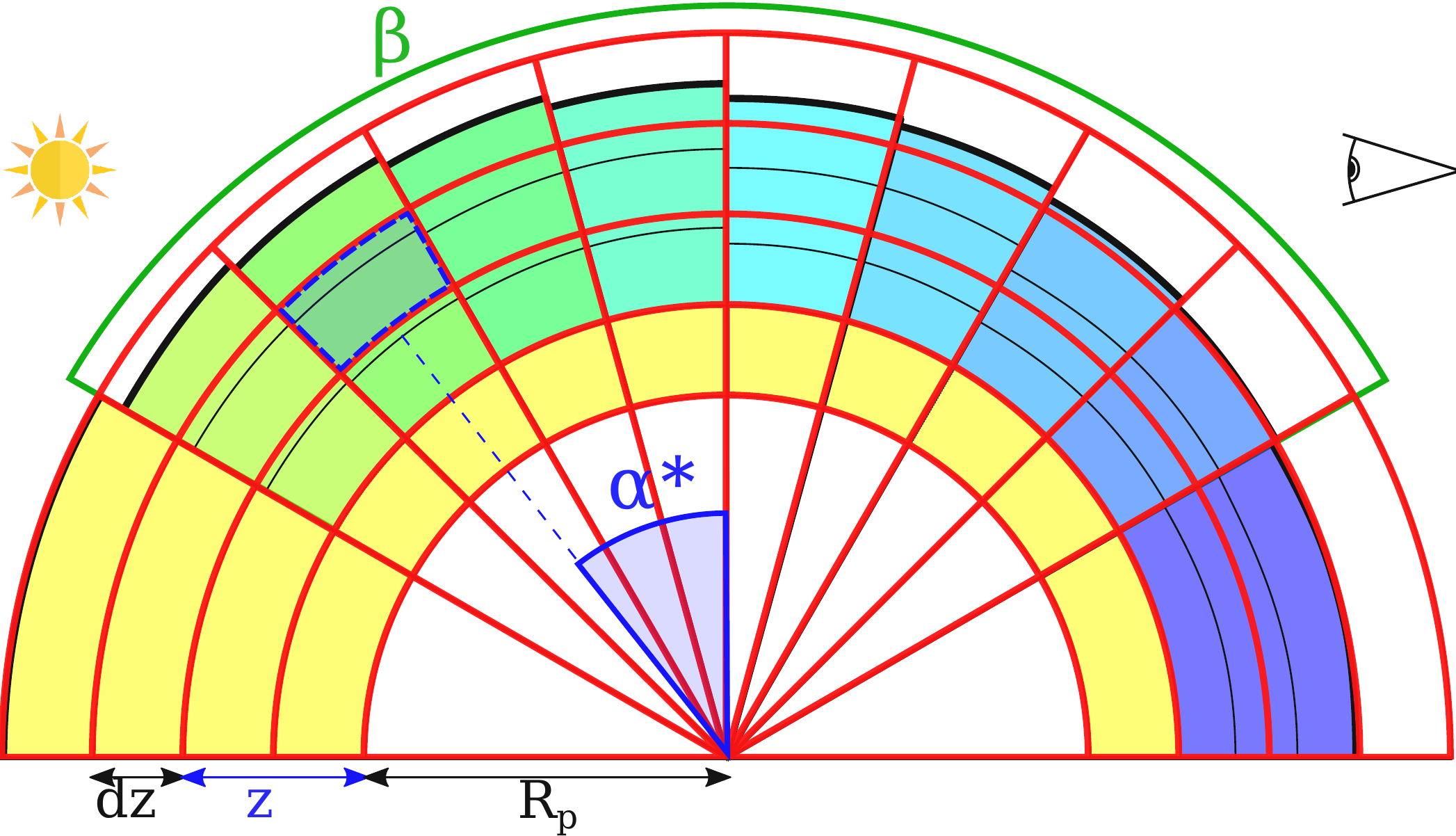}
	\caption{The red-colored grid is a polar grid $\tz$ of which the radial ticks are set as the altitude corresponding to levels of pressure in the day side of the planet. The input pressure-based grid $\tp$ is also represented in the same coordinate system (colored cells from yellow to purple).}
	\label{planet_P_z}
\end{figure}

A 1D pressure profile is first defined within the assumption of a plane-parallel atmosphere, in a fashion similar to \taurex \citep{Al-Refaie2021}.
The temperature is then defined in a 2D coordinate system $\tp$ of which one dimension is angular and the second is pressure-based.
This coordinate system is shown by the underlying (black) grid in \cref{planet_P_z}.
The temperature, defined by Eq.~\ref{eq:temperature_P}, follows a linear transition between $-\beta/2$ and $\beta/2$ as proposed in \citet{caldas2019effects, Wardenier2022}.
\setlength\arraycolsep{1pt}
\begin{equation}
\label{eq:temperature_P}
\left\{ \begin{array}{lll}
	P > P_{iso}, & T = & T_{deep},\\
	\multirow{3}{*}{$ 
	P < P_{iso},
	\left\{ \begin{array}{l}
		2\al \geq \beta , \\
		2\al \leq -\beta, \\
		-\beta < 2\al < \beta,
	\end{array} \right. %
	$} 
	& T = & T_{day},\\
	& T = & T_{night},\\
	& T = & T_{night} \\
	& & + (T_{day} - T_{night}) 
	\frac{\al+\beta/2}{\beta}.
\end{array} \right.
\end{equation}
It relies on three temperature variables ($T_{day}$, $T_{night}$, $T_{deep}$), an angle parameter ($\beta$) and a pressure level defining the upper limit of an isothermal annulus ($P_{iso}$). This parametrization, although simplistic, is based on global climate model (GCM) simulations \citep{Showman2015,Parmentier2018,tan2019} to capture the most salient features of the thermal structure of HJs. 

The temperature is shown by the colors (yellow to purple) in \cref{planet_P_z}.
The data (pressure, temperature, abundances) in the input grid $\tp$ is then interpolated in an altitude-based coordinate system $\tz$, shown by the red grid in \cref{planet_P_z}.
Using the new coordinate system $\tz$, rays of light crossing the atmosphere may be subdivided into a list of segments, as illustrated in \citet{falco2020taurex2D}, using simple geometric formula. There is one ray per layer in the model. The length of a ray $r$ which crosses the $i$-th atmospheric layer is identified with $\Dl_{\rc,i}$. The optical depth can then be computed for each ray $r$ with:

\begin{equation}
        \tr = \sum_{\substack{i}}
        \frac{P_{\rc,i}}{k_B T_{\rc,i}}
        \left(\sum_{m=1}^{N_{gas}}
        \chi_{m,\rc,i} \sigma_{m, \lambda} +
        \sum_{j=1}^{N_{con}} k_{mie, j}
        \right)\Dl_{\rc,i},
        \label{eq:tau_layer}
\end{equation}
where $P_{\rc,i}$ and $T_{\rc,i}$ are the pressure and temperatures of the cell corresponding to the segment $i$ of the ray $\rc$, $k_B$ is the Boltzmann constant, $\chi_{m,\rc,i}$ is the volume-mixing ratio of the $m$-th molecule, $\sigma_{m, \lambda}$ is the total cross-section of Rayleigh scattering and molecular and continuum absorptions, and finally, $k_{mie, j}$ is the cross section associated with the Mie scattering for the $j$-th aerosol. 
$N_{gas}$ is the number of chemical species, $N_{con}$ is the number of aerosols, We take into account collision-induced absorption for H$_2$-H$_2$ and H$_2$-He pairs from HITRAN \citep{richard2012}. Formally, the contribution of CIA opacities is normalized to the number of H$_2$ molecules and included in $\sigma_{\mathrm{H}_2, \lambda}$.

The volume mixing ratios for all the species can also be differentiated with a day and a night value arrays. In this work, we have set the number of layers to 100 and the number of angular points (or slices) to 24. These two numbers are a good compromise between precision of the model and computational speed \citep{falco2020taurex2D}.
Two angular points correspond to the full day and night sides, while the rest equally subdivide the space between $-\beta/2$ and $\beta/2$, accounting for the linear transition in-between.

\subsection{Parameters}
\label{parameters}

\begin{table*}[!htbp]
	\centering
	\caption{Input parameters for the three atmosphere configurations.\label{tab:val_input}}
	\begin{tabular}{|c|c|c|c|c|c|c|c|}
		\hline
		Test & $T_{\text{deep}}$  & $T_{\text{up, day side}}$ & $T_{\text{up, night side}}$ & $P_{\text{iso}}$ & log$_{10}$[H$_2$O]$^*$ & log$_{10}$[CO] & $\beta$\\
		\hline
		1D Fully Isothermal  &  2500K & 2500K & 2500K & $10^4$Pa & -3.30 & -3.36 & - \\
		\hline
		Two Layers &  1800K & 2600K & 2600K & $10^4$Pa & -3.30 & -3.36 & - \\
		\hline
		Day/Night & 2500K & 3300K & 500K & $10^4$Pa & -3.30 & -3.36 & 20$^\circ$\\
		\hline
	\end{tabular}
	\\
	$^*$These values change as a function of pressure and temperature when water dissociation is assumed.
\end{table*}

We tested \taurex~2D retrievals using three different configurations described in Tab \ref{tab:val_input}.
In all three configurations the thermal structure of the atmosphere is described using the following free parameters:

\begin{enumerate}
    \item $\beta$ angle: the angle across the terminator line within which we compute the radiative transfer equation;
s	\item $P_{iso}$: the pressure of the boundary between the lower and the upper level;
	\item $R_p$: the 10\,bar radius of the planet;
	\item $T_{day/night/deep}$: the temperature in the day and night side of the planet and in the deeper isothermal region, respectively;
	\item $P_{clouds}$: the pressure of the upper level of the cloud deck.
\end{enumerate}

The \taurex~2D retrieval code is configured to have a deep atmospheric region  where the molecular abundances and the temperature remain constant: at pressures higher than $P_{iso}$, the temperature is equal $T_{deep}$ and the volume mixing ratio of each molecule has a constant value. 
In the upper part of the atmosphere, where $P < P_{iso}$, the temperature is different between the day side ($T_{day}$) and the night side ($T_{night}$).

For the chemistry, we can use two different schemes:
\begin{enumerate}
\item Constant (or ``free'') chemistry mode, where volume mixing ratios are constant within each column ($\alpha$) for $P < P_{iso}$. In this mode, \taurex~2D can fit independently the abundances on each hemisphere, this implies three free parameters per molecule ($[mol]_{side1}$, $[mol]_{side2}$ and $[mol]_{deep}$).
\item Thermal dissociation mode, where we assume local equilibrium chemistry. In this mode, the free parameters are the deep abundances of each molecule ($[mol]$), that are considered uniform over the planet but we recompute the local abundances as a function of the local pressure and temperature following \citet{Parmentier2018}. In the regime we consider, these changes in abundances are mainly driven by thermal dissociation. 
\end{enumerate}

The thermal dissociation module we used in this work follows the parametrization of \citet{Parmentier2018}, which uses a modified version of the NASA CEA Gibbs minimization code \citep{gordon1984} in a grid of models to investigate gas and condensate equilibrium chemistry in differrent atmospheric conditions in substellar objects \citep{moses2013,kataria2015,skemer2016,wakeford2017,burningham2017,marley2017}.

While the constant chemistry mode could seem better as it offers more flexibility, we will show that its higher number of parameters creates more degeneracies than the other parametrization (see Section \ref{symmetry}) so that the thermal dissociation mode will be our default except when stated otherwise. 

In the tests we show in the next sections, we assumed a constant 30 ppm error bars in the considered wavelength range $0.5\mu$m-$10\mu$m. The assumed error was kept throughout the whole spectral range with a normal distribution \citep{greene2016}. This value was used as a noise floor in \citet{Pluriel2022,pluriel2020a} and it is a good representation of the average noise computed using the estimated mean number of photoelectrons in \citet{cowan2015} for a WASP-121b like star at 270 pc during a single transit. This wavelength range will be covered using MIRI and NIRSpec instruments on board of the JWST space telescope \citep{jwst2006}.

\section{Test and Validation}
In all the test cases, we assume the same planetary and stellar parameters, that of the system WASP-121, whose values are specified in Tab \ref{tab:system_input}.

\begin{table}[!htbp]
	\centering
	\caption{Input planetary and stellar parameters.\label{tab:system_input}}
	\begin{tabular}{ccc}
    \toprule
    \toprule
    \multicolumn{3}{c}{Planetary parameters} \\
    \cmidrule{1-3}
    $R_p$ & \multicolumn{2}{c}{$1.807R_{Jupiter}$}  \\
    $M_p$ & \multicolumn{2}{c}{$1.183M_{Jupiter}$}  \\
    $g_s$ & \multicolumn{2}{c}{$9.39ms^{-2}$} \\
    $H_e/H_2$ & \multicolumn{2}{c}{0.25884514} \\
    $[\log{P_{min}},\log{ P_{max}}] $ & \multicolumn{2}{c}{$[-4, 6] (Pa)$} \\
    \cmidrule{1-3}
    \multicolumn{3}{c}{Stellar Parameters} \\
    \cmidrule{1-3}
    $R_*$ & \multicolumn{2}{c}{ $1.458R_{\odot}$ } \\
    $T_{eff}$ & \multicolumn{2}{c}{ $6460K$ } \\
    \cmidrule{1-3}
    \multicolumn{3}{c}{Retrieval Bounds} \\
    \cmidrule{1-3}
    $\beta$ (degrees) & \multicolumn{2}{c}{ $0 - 30$ } \\
    $T$ (K) & \multicolumn{2}{c}{ $100 - 3600$ } \\
    $\log{P}_{iso}$ (Pa) & \multicolumn{2}{c}{ -4 $-$ 6 } \\
    $R_p$ ($R_{\textnormal{jupiter}}$) & \multicolumn{2}{c}{ 0.9 $-$ 2.7 } \\
    $\log$ mixing ratios & \multicolumn{2}{c}{ -10 $-$-1 } \\
    $\log{P}_{clouds}$ (Pa) & \multicolumn{2}{c}{ -4 $-$ 6 } \\
    \bottomrule
\end{tabular}
\end{table}

\taurex~2D works within the same Bayesian framework as the one dimensional \taurex retrieval code \citep{Al-Refaie2021}. The atmospheric retrieval which uses a two-dimensional forward model is more computationally intensive than the one calculated with a one-dimensional forward model, due to its additional number of parameters and geometrical dimensions.

\begin{figure}[!htbp]
	\centering
	\noindent\includegraphics[width=0.5\textwidth]{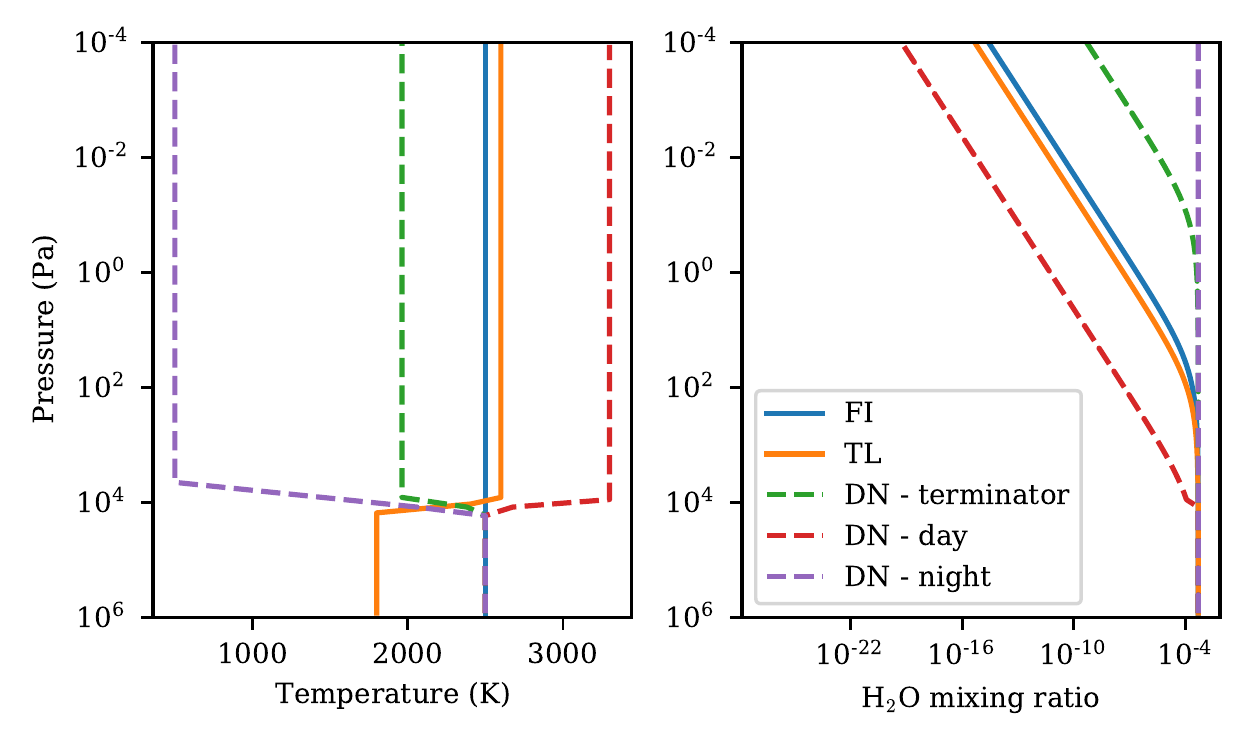}
	\caption{Pressure Temperature (\textbf{left}) and $\log{[\text{H}_2\text{O}]}$ mixing ratio (\textbf{right}) profiles for the three different test cases: \textbf{FI}: Fully Isothermal, \textbf{TL}: Two Layers and \textbf{DN}: Day Night cases. The DN case has been split into three curves showing the profiles at the therminator, the day and the night side. \label{fig:input_PT}}
\end{figure}

Before focusing on a realistic test case, computed using SPARC/MITgcm \citep{showman2009}, it is fundamental to test the reliability of the \taurex~2D code on some synthetic spectra generated by itself. We decided to test the code using the three cases described in Tab \ref{tab:val_input}. 
For these tests, the input spectra have been generated by the \taurex~2D forward model itself.  The three cases represent three different scenarios with different PT profiles and, then H$_2$O mixing ratio profiles as shown in Fig. \ref{fig:input_PT}.  All the simulated spectra have been generated assuming a resolution $R=\lambda_0/\Delta\lambda = 100$ where $\lambda_0=5.25\mu m$ and $\Delta\lambda = [0.5\mu m - 10\mu m]$.

The retrieval tests were performed using the Multinest algorithm \citep{feroz2009} with 1000 live points and uniform prior distribution between the bounds shown in Tab \ref{tab:system_input}.
In the next three subsections we discuss the results for the three test configurations. For the sake of conciseness, the fitted spectrum and posterior distributions will be shown only for the third, most complex case.
To summarize, the three tests 
show that \taurex~2D can perfectly retrieve itself when there is enough information in the input spectrum. 

\subsection{Fully Isothermal atmosphere}
A fully isothermal atmosphere represents the simplest assumption in our test set. We assumed a constant vertical and day/night composition with H$_2$O and CO defined in Tab \ref{tab:val_input}.

The solution found is consistent with the input model. \taurex~2D found a solution which is consistent with the input parameters. \taurex found a volume mixing ratio for $\log{[H_2O]} = -3.30^{+0.01}_{-0.01}$ for water and $\log{[CO]} = -3.36^{+0.05}_{-0.06}$ and for CO, centered around the ground truth for this case. Since we used a forward model with two vertical layers, divided at pressure $P_{iso}$, \taurex found the best model where the upper layer is suppressed at the top of the atmosphere around $10^{-3.07}$Pa and treated the rest of the atmosphere as isothermal with $T=2502^{+10}_{-11}$, consistently with the input model.

The 2D model of \taurex is equivalent to the 1D version of \taurex in the case of isothermal atmospheres, as demonstrated in \citet{falco2020taurex2D}. Therefore, this retrieval test is a confirmation of the consistency of the implementation with its 1D predecessor.

\subsection{Two Vertical Layers atmosphere}

The test case with two layers is an atmosphere where the lower layer has a temperature $T_{deep}=1800K$ and the upper temperature has $T_{upper, day/night}=2600K$ but we do not have any day/night difference.
We use the deep abundances defined in Tab \ref{tab:val_input} and let molecules dissociate as a function of the pressure and temperature.
This test is important to verify whether our model can find the input vertical structure of the atmosphere, without including day/night effects. When $T_{upper} \neq T_{deep}$, the code should recover the input $P_{iso}$ and two different temperatures. 
When $T_{upper} = T_{deep}$, it should converge to the isothermal configuration. 

We chose these parameters for the two-layers configuration to demonstrate how \taurex~2D can converge to the correct temperature structure, given that the composition remains the same in both hemispheres and with a temperature inversion between the lower and the upper part of the atmosphere. For such a test, we put the pressure threshold $P_{iso}$ between the two layers at 1 mbar, which corresponds to the range of pressures probed in such hot atmospheres, thus ensuring that the input spectrum should encode some information about the temperature change.  

\taurex~2D found a $T_{side1}=2500^{+700}_{-1400}$, $T_{side2}=2600^{+600}_{-1400}$, $T_{deep}=2594^{+8}_{-12}$ and $\log{P_{iso}}=-2.89^{+0.82}_{-0.51}$. 
\taurex favors a thin upper atmosphere and fits the lower layer with an isothermal at $\sim2600 K$, which is the input upper temperature. 
This means that we managed to probe only the upper layer of the atmosphere. 
The retrieved molecular abundances found are compatible with the input values $\log{[H_2O]} = -3.33^{+0.01}_{-0.01}$ for water and $\log{[CO]} = -3.34^{+0.05}_{-0.04}$.

\subsection{Day/night atmosphere}
\label{daynight}

\begin{figure}[!htbp]
	\centering
	\noindent\includegraphics[width=\columnwidth]{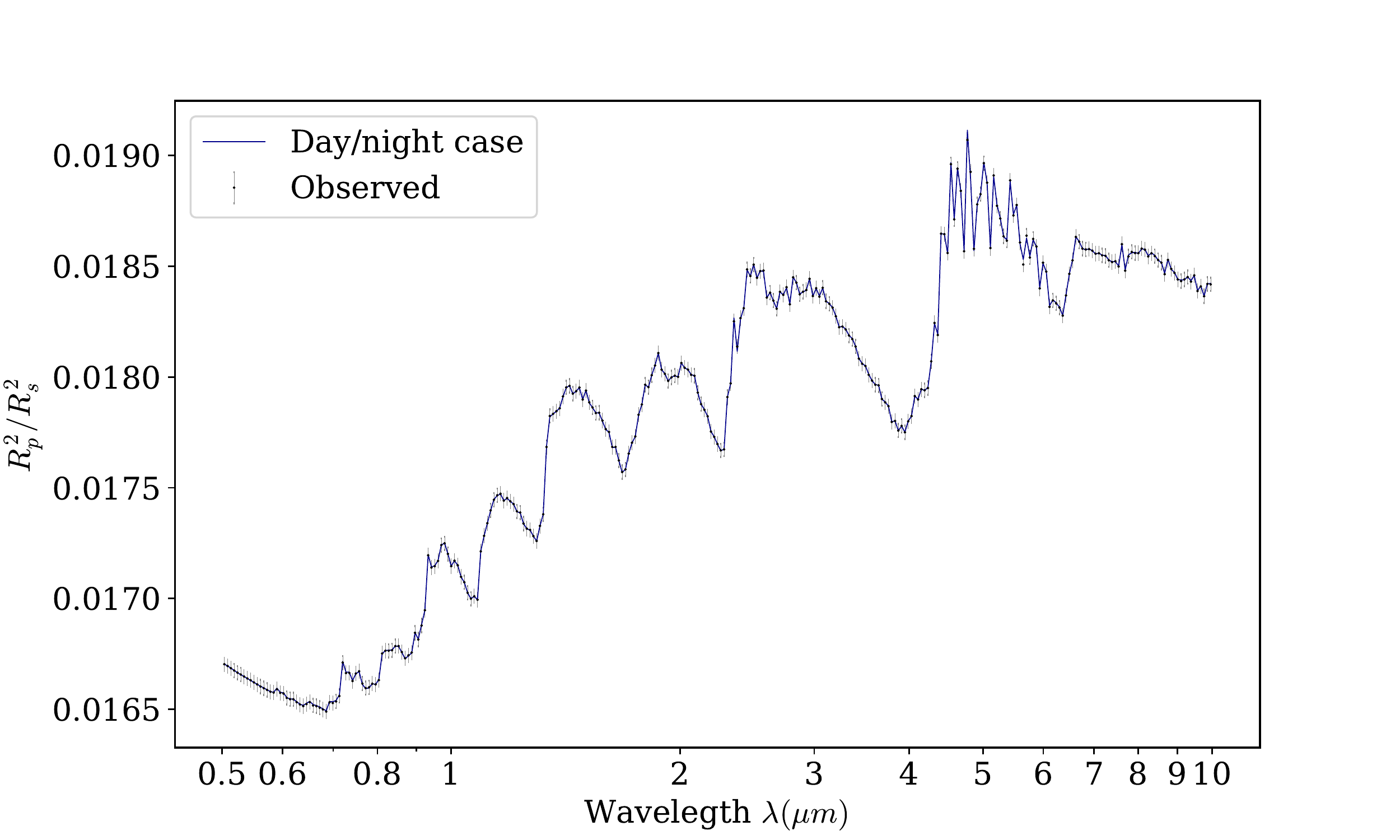}
	\caption{Best fit spectrum for the Day/Night convergence test. \label{fig:DNspectrum}}
\end{figure}

With this complex test, we explore the most complex atmospheric configuration in our test cases. Indeed, the input spectrum was generated assuming an isothermal inner annulus at $T_{deep}=2500K$ a day side temperature $T_{day}=3300K$ and a night side of $T_{night}=500K$ with a $\beta$ angle of  20$^\circ$ (see Tab \ref{tab:val_input} for the exact setting). 
The day/night thermal difference is realistic according to GCM models of this type of planet (e.g. WASP-121b) and, within this scenario, the $\beta$ parameter can give us a hint of the 3D structure of the planet, since within this angle we probe both, the day side and the night side, around the terminator line. 
Based on this GCM equatorial temperature maps, the atmosphere can be approximated to an isothermal day and night sides, with a sharp transition region between the two sides (represented by the $\beta$ parameter), and a isothermal ring under a certain pressure. Within this scenario, the parameter can give us a hint as to the 3D structure of the planet, since within this angle we probe both, the day side and the night side, around the terminator line. In Fig. \ref{fig:DNspectrum} we show the best fit model for this last test case.

\begin{figure*}[!htbp]
	\centering
	\noindent\includegraphics[width=\textwidth]{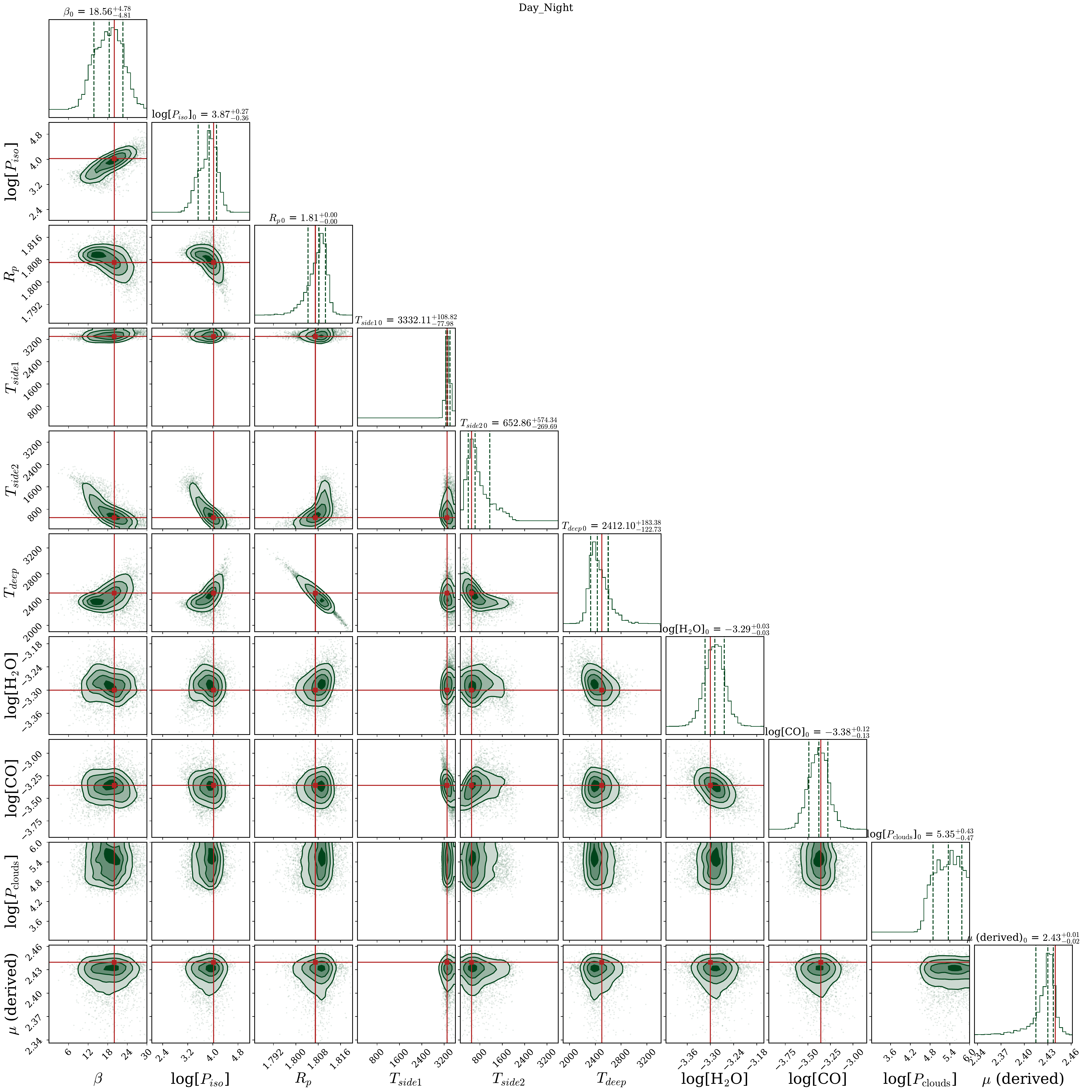}
	\caption{Posterior distribution for the Day/Night case. The input spectrum refers to an hot atmosphere with three different equilibrium temperatures: lower temperatures, upper day and upper night. As in the previous models, H$_2$O and H$_2$ can dissociate as a function of P (Pa) and T (K). The red lines represent the input values to generate the input spectrum.  \label{fig:DNposterior}}
\end{figure*}

In Fig. \ref{fig:DNposterior} we can see how \taurex finds a unique solution around the ground truth represented by the red vertical lines. 

With this final convergence test, we can see the reliability of \taurex~2D with handling spectra from bidimensional models. We let the $P_{iso}$ and $\beta$ as free parameters and, we give \taurex the possibility to look for the best spectrum. 

The $\beta$ parameter retrieved with \taurex2D, in particular, would give us valuable information to model three dimensional effects of an exoplanetary atmospheres. We see in Fig. \ref{fig:DNposterior} that \taurex found the parameters of the input spectrum.

\section{Intrinsic degeneracies in transmission spectra: can we recover independently the compositions of the day and night sides?}
\label{symmetry}

By construction, the transmission spectrum of a planet only depends on the \textit{total} optical depth of the atmosphere along the various rays of light passing through the limbs (see Eq.\,\ref{eq:tau_layer}). Therefore, it can be seen right away that the transmission spectrum of a planet would remain unaffected if we took a mirror image of the planet, putting the day side away from the star. 

This intrinsic symmetry of the problem tells us that, \textit{technically}, a retrieval algorithm cannot determine the conditions on the day and night sides. At best, it can infer the presence of different conditions on both hemispheres and we have to use our physical insight to decipher what conditions correspond to what side; for example be saying that the hottest hemisphere is probably the dayside one. This is why we use $T_{side1}$ and $T_{side2}$ as independent parameters in our retrievals.

But the fact that we measure only the total optical depth creates another degeneracy: to first order, only the total column density of a species along the ray matters. To take an extreme example, if we had a completely isothermal planet with two hemispheres having different chemical compositions, the effect of a trace molecule (mol) on the spectrum would be unaffected as long as
\begin{equation}
  [\mathrm{mol}]_{day} + [\mathrm{mol}]_{night} = cst.
  \label{eq:linera_combination}
\end{equation}
A homogeneous atmosphere would have the same signature as an atmosphere where all the molecules would be in a single hemisphere. 

\begin{figure*}[!htbp]
	\centering
	\noindent\includegraphics[width=\textwidth]{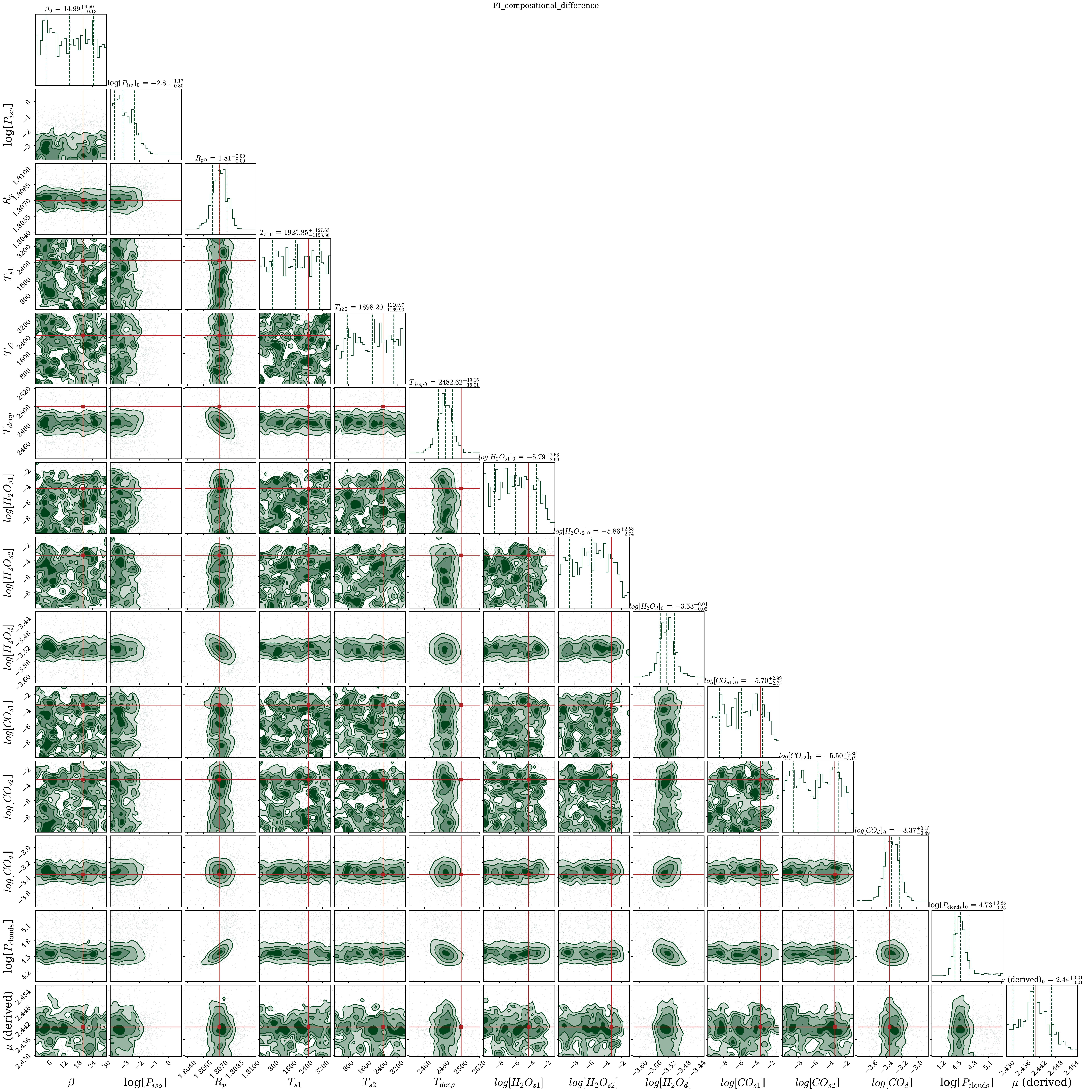}
	\caption{Posterior distribution for a particular case of the Fully isothermal atmosphere with different water abundances in the day ($\log[H_2O]_{day} = -5.3$) and the night side (($\log[H_2O]_{night} = -4.3$)) \label{fig:FI_comp_diff}}
\end{figure*}

To check the performance of \taurex~2D with a fully isothermal atmosphere and a compositional difference between the day side and the night side, we simulated an input synthetic spectrum as in the fully isothermal test case, with the difference that the water abundance changes from the day ($\log[H_2O]_{day} = -4.3$) to the night side ($\log[H_2O]_{night} = -3.3$). As shown in the posterior distributions (Fig.  \ref{fig:FI_comp_diff}), \taurex~2D favors the solution in which the upper layer is flattened towards the upper edge of the planetary atmosphere, in a small annulus with a pressure lower than $\log[P](Pa)= -2.81^{+1.17}_{-0.80}$ and fits an isothermal atmosphere using only the lower layer. The input temperature is found within 2-sigma and the lower water abundance $\log[H_2O]_{deep} = -3.53^{+0.04}_{-0.05}$ is within 1-sigma from the arithmetic average between the input abundances $\log[H_2O]_{deep, input} = -3.56$.

Of course, having different temperatures on both sides formally breaks this degeneracy, because of the scale height effect as well as the temperature dependence of the molecular opacities \citep{MacDonald2021trident}. But the question remains whether there is enough information in a transmission spectrum to retrieve the composition of both sides independently. 

\begin{figure*}[!htbp]
\centering
\includegraphics[width=\textwidth]{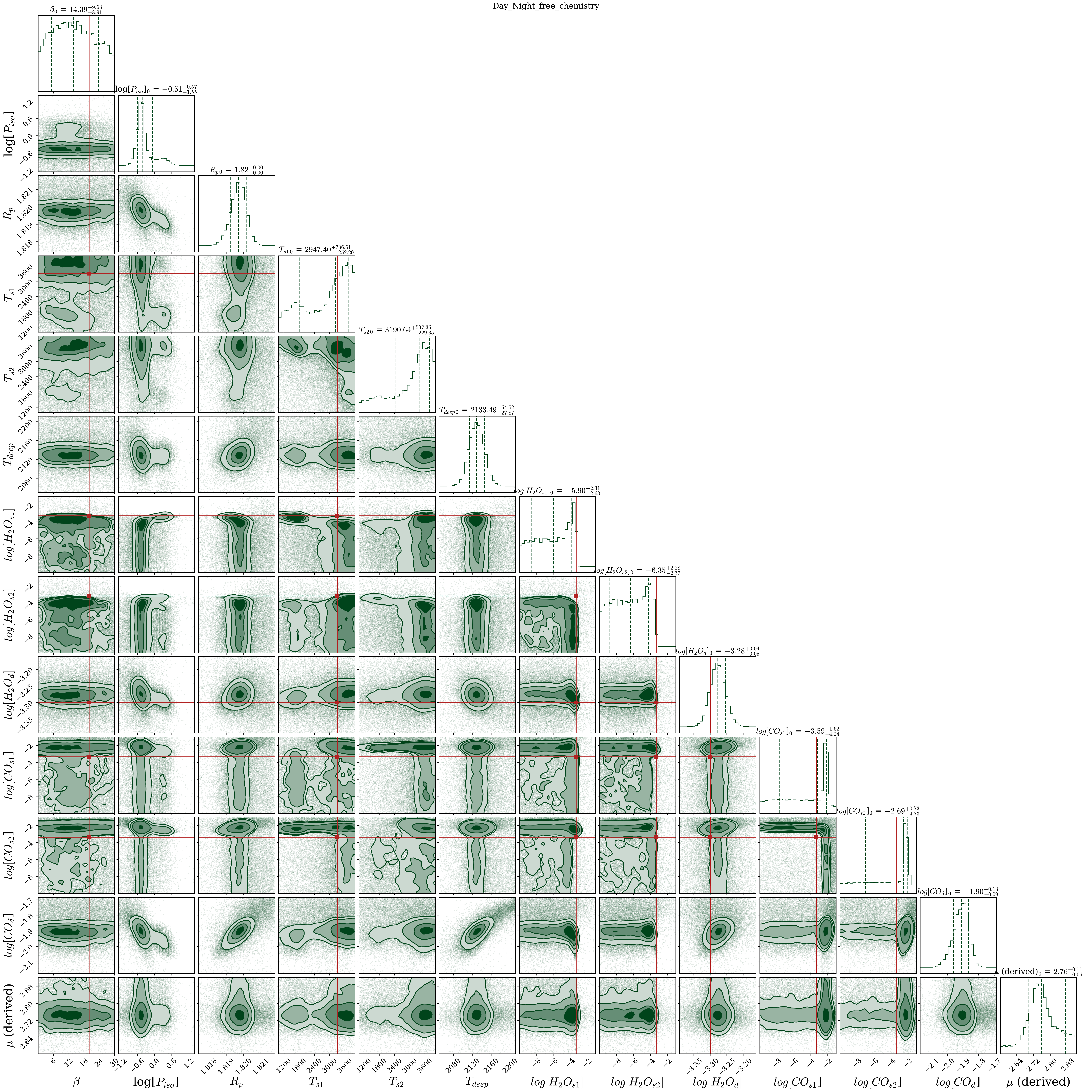}
\caption{Posterior distribution for a retrieval test made using the same ''Day/Night'' input spectrum as in Fig. \ref{fig:DNposterior} but retrieved using a forward atmospheric model with independent compositions for the day and night sides. One can see the strong cross-correlations between the abundances of a given molecule on the day and night sides.
\label{fig:independent_posteriors}}
\end{figure*}

\begin{figure}[!htbp]
\centering
\includegraphics[width=\columnwidth]{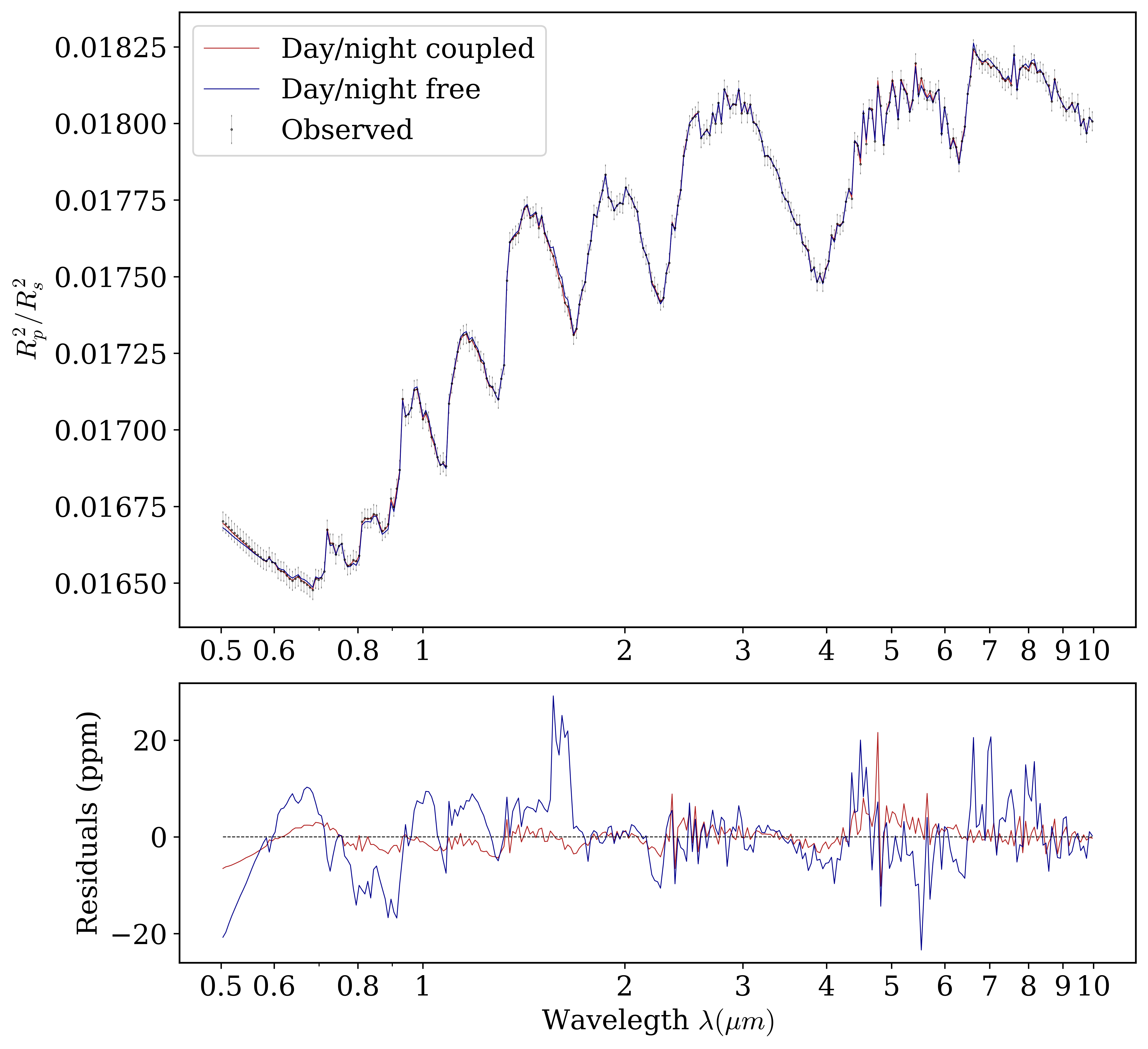}
\caption{Spectral comparison between the two Day/Night retrieval cases.  We can \textbf{see} that both the coupled model (the model that considers thermal dissociation in the atmosphere) and the ``free'' model (which is the one that considers a constant chemistry, but independent on both sides of the planets) lead to a consistent spectrum. \label{fig:comparison_daynight_spectra}}
\end{figure}

To test this, we performed another retrieval on the `Day/night'' input spectrum used in sect.\,\ref{daynight} using \taurex~2D\textbf{'s} ability to model the planet with independent compositions for the day and night sides as described in Section \ref{parameters} (''free'' mode). The posterior distributions for this retrieval are shown in Fig. \ref{fig:independent_posteriors}, and the best fit spectra found by the two models (along with the residuals) are shown in Fig. \ref{fig:comparison_daynight_spectra}.

As can be seen in the bottom panel of Fig. \ref{fig:comparison_daynight_spectra}, the residuals are well below the 30\,ppm noise floor used as uncertainty in these retrievals. The two approaches thus give an acceptable fit. However, it can be seen in Fig. \ref{fig:independent_posteriors} that the posterior distributions of the ''free'' retrieval show a lot of cross-correlation between several parameters. In particular, as can be expected from the argument above, when the CO or H$_2$O abundances are high on one side, it can be very small on the other. This model also seems to miss the temperature dichotomy between the two hemispheres of the planet. 

To see whether these two forward models can be statistically differentiated,
we introduce the (logarithmic) Bayes factor:

\begin{equation}
  \mathscr{B} = \log{\frac{E_\mathrm{d}}{E_\mathrm{f}}} = \log{E_\mathrm{d}} - \log{E_\mathrm{f}},
  \label{eq:logbayes_factor}
\end{equation}

where $E_\mathrm{d}$ and $E_\mathrm{f}$ are the Bayesian evidence for the retrievals with dissociation (d) and free (f) models.

Given that the Bayesian evidence of the retrieval which uses thermal dissociation and coupled day/night sides is $E_\mathrm{d}=2914$ while the one related to the model which uses independent day/night sides is $E_\mathrm{f}=2908$, it follows that the Bayes factor is $\sim$6 giving a slight but significant advantage to the coupled model. In this case, it seems that the higher number of free parameters penalizes the model with ''free'' composition.

This seems to show that there is not enough information in the input spectrum (at least with the level of uncertainty expected for a JWST like observation) to retrieve simultaneously and independently the composition of both sides of the planet. Going further, the results even imply that the information on the temperature dichotomy in a transmission spectrum mainly comes from the link between temperature and composition in the observed atmosphere. Assuming chemical equilibrium brings a considerable amount of prior knowledge on the atmosphere. If this link seems robust in Ultra hot Jupiters where molecular dissociation is fast enough to happen close to chemical equilibrium, it should be revisited for cooler atmospheres. For this reason, the ability to retrieve the temperature dichotomy on cooler planets might be greatly reduced compared to what is shown in \citet{lacy2020}, where they assumed chemical equilibrium to hold down to dayside temperatures near 1200\,K. This issue should be thoroughly investigated.

For all these reasons, in the following, we will keep using the model assuming thermal dissociation at chemical equilibrium as it possesses less free parameters and results in less correlated posterior distributions between the parameters. 
We should however keep in mind that this assumption should be revisited when modelling cooler atmospheres.

\section{A GCM retrieval}
\label{GCMretrieval}

After validating the \taurex~2D retrieval code on some self-generated spectra, we now test its abilities on a more realistic atmosphere, generated by the SPARC/MIT GCM and described in detail in \citet{Parmentier2018}. We simulated the transmission spectra using 30ppm photon noise as described in section \ref{parameters}. We used the temperature map obtained by the GCM simulation and compute the radiative transfer equation through a hot Jupiter atmosphere with \pytmo 2.0 \citep{falco2020taurex2D}. 
The representative planet for this test is a WASP-121~b model, where molecular species can thermally dissociate as a function of the pressure and temperature. 
In this model, we include TiO and VO in the atmosphere as they have been detected on WASP-121\,b by \citet{Evans2018}. These molecules are optical absorbers that heat the day side of the planet and create a strong thermal inversion \citep{Fortney2008,Parmentier2015,Parmentier2018}. Other molecules, ions and metals (such as Iron detected in the UHJ WASP-76 b by \citet{Ehrenreich2020}) also contribute significantly to thermal inversion in UHJs, as pointed out by \citet{Lothringer2018}. Based on this GCM model of WASP-121 b, we simulate two spectra from two different atmospheric configurations to test its impact on the retrieval analysis:

\begin{itemize}
    \item A solar metallicity atmosphere with H$_2$O and CO as trace gases;
    \item A solar metallicity atmosphere with H$_2$O, CO, TiO and VO as trace gases.
\end{itemize}

The presence of TiO and VO covers indeed some fundamental wavelengths where Rayleigh scattering otherwise dominates. In the absence of clouds, this wavelength range allows the model to calibrate the pressure in the atmosphere and thus to have better absolute abundances. We thus want to test whether hiding these regions leads to additional biases on the retrieved parameters.

A crucial parameter related to the three-dimensionality of the planet is the $\beta$ parameter. The $\beta$ angle is a 2D parameter and therefore its equivalent for a 3D GCM simulation is hard to define. 
For the following input spectra, the best approximation we can make is $5_{-3}^{+3}$ degrees based on the GCM equatorial temperature map, indicating a very sharp transition between the day and the night sides.

\subsection{Atmosphere without optical absorbers}
\begin{figure}[!htbp]
	\centering
	\noindent\includegraphics[width=\columnwidth]{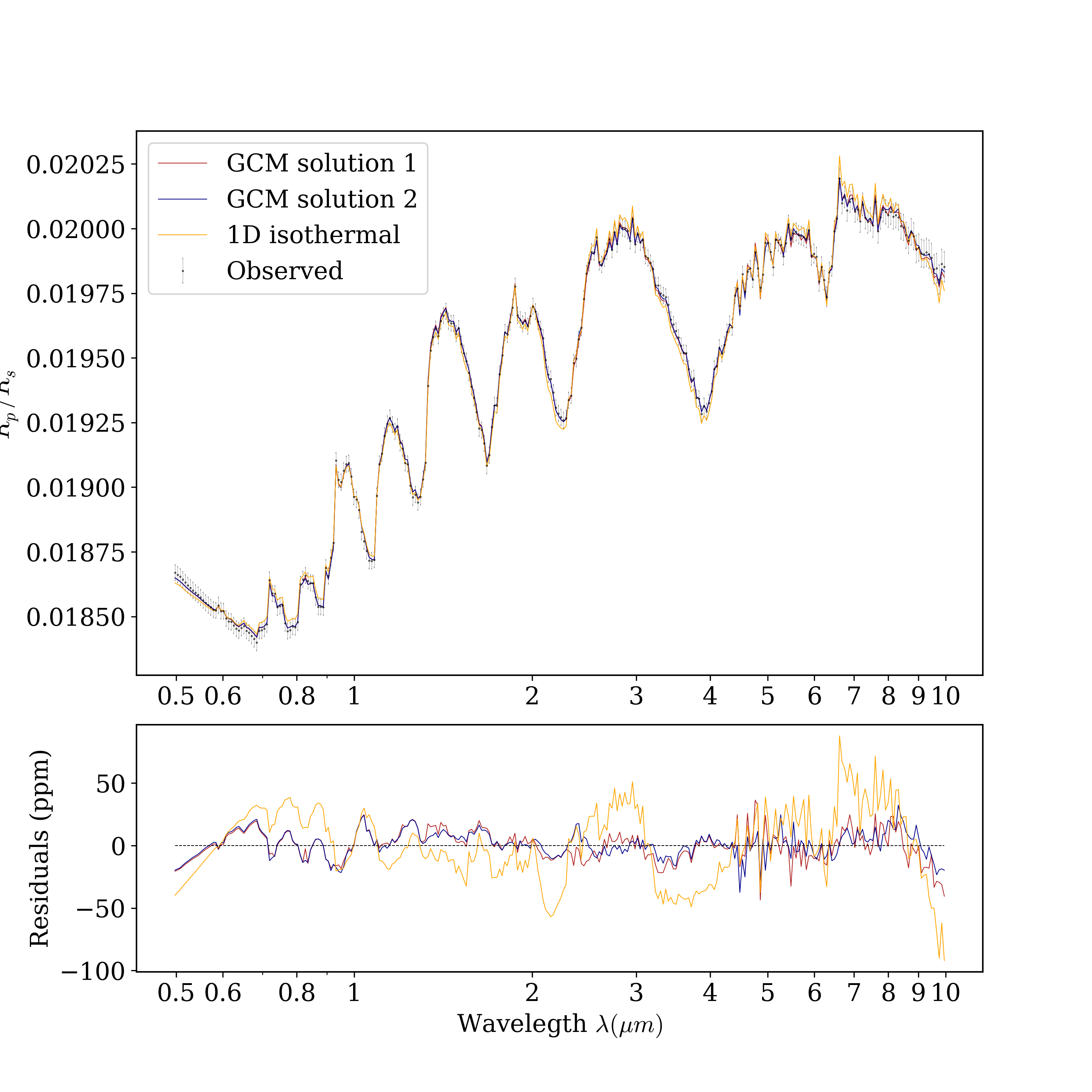}
	\caption{Best fit model for an atmosphere without optical absorbers. \taurex~2D retrieval code found two possible solutions for this case. The comparison with the retrieval made with a one-dimensional isothermal forward model shows that \taurex~2D leads to a better interpretation of the atmosphere. \label{fig:GCMnoabs_spectrum}}
\end{figure}

We start analyzing an atmosphere without the presence of optical absorbers (e.g. molecules like TiO and VO). The retrieval performed by \taurex~2D converged to the best-fit solution shown in Fig. \ref{fig:GCMnoabs_spectrum}.
In this test we can also see a degenerate solution converging to two different $\beta$ angles and temperatures for the two sides of the planet.

\begin{figure*}[!htbp]
	\centering
	\noindent\includegraphics[width=\textwidth]{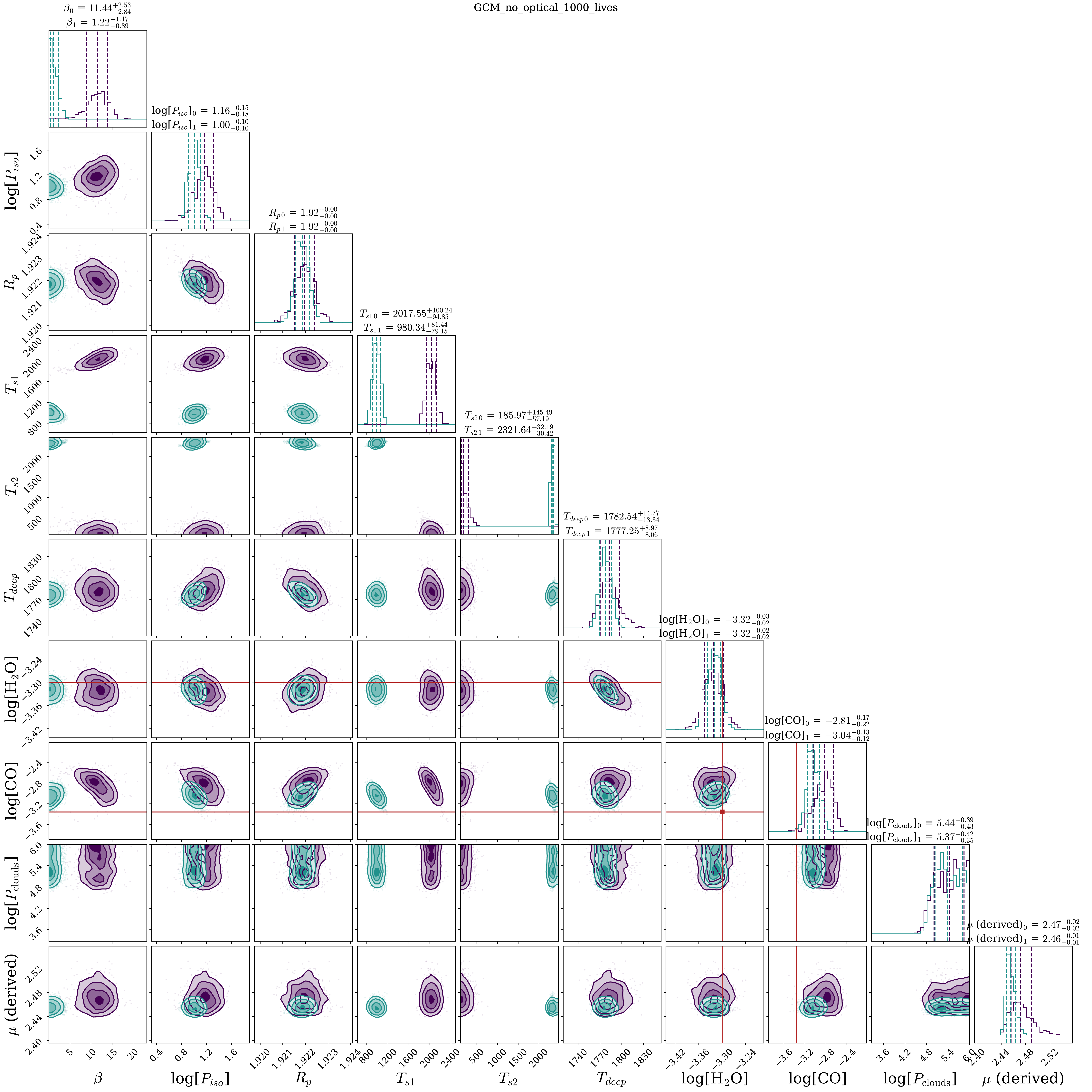}
	\caption{Posterior distribution for the case where we used as input a WASP-121b GCM's synthetic atmosphere using only H$_2$O and CO as trace gases. The red lines correspond to the input deep abundances from the GCM model atmosphere.\label{fig:GCMnoabs_posteriors}}
\end{figure*}

The two solutions found by \taurex have different physical properties and the best fit parameters are shown in Fig. \ref{fig:GCMnoabs_posteriors}.

\taurex~2D managed to converge around a realistic solution which is compatible with the ground truth, i.e. the parameters used to generate the input spectrum. Indeed, as shown by the red line in Fig. \ref{fig:GCMnoabs_posteriors}, the deep atmosphere trace gas parameter volume mixing ratios are compatible with the ground truth of the input model within $1\sigma$ error bars for both solutions. In addition, the temperatures obtained in both solutions reflects realistically the ones of WASP-121b. This two-dimensional model is more realistic and accurate than a simpler one-dimensional model, since it is able to catch day-night differences that, as shown in \citet{pluriel2020a}, are crucial in Hot Jupiters like WASP-121b.

We find $\beta_0 = 1.2_{-0.9}^{+1.2}$ and $\beta_1 = 11.4_{-2.8}^{+2.5}$, respectively, the $\beta$ angle for first and second solution, whose measures are correlated with the temperature values found in the two sides, respectively $T_{night} = 190_{-60}^{+150}$ and $T_{day} = 2000_{-100}^{+100}$ for the first solution, $T_{night} = 980_{-80}^{+80}$ and $T_{day} = 2320_{-30}^{+30}$ for the second solution. Interestingly, both solutions give the same pressure level and the same temperature for the deep layer ($P_{iso}\sim10$ Pa and $T_{deep}\sim1780K$). It means that the retrieval can probe approximately up to that atmospheric region, after which we approximate the atmosphere with an isothermal annulus of $T_{deep}\sim1780K$.
The $\beta$ parameter gives us a hint of the three-dimensional structure of the planet, since it spans above the day side and the night side, and a dramatic change on this parameter can lead to a dramatic change in the retrieved temperature on both sides of the planet. In both solutions, the transition between the day and the night side retrieved is sharp since it is less than $12^{\circ}$. 

\begin{figure}[!htbp]\centering
	\noindent\includegraphics[width=\columnwidth]{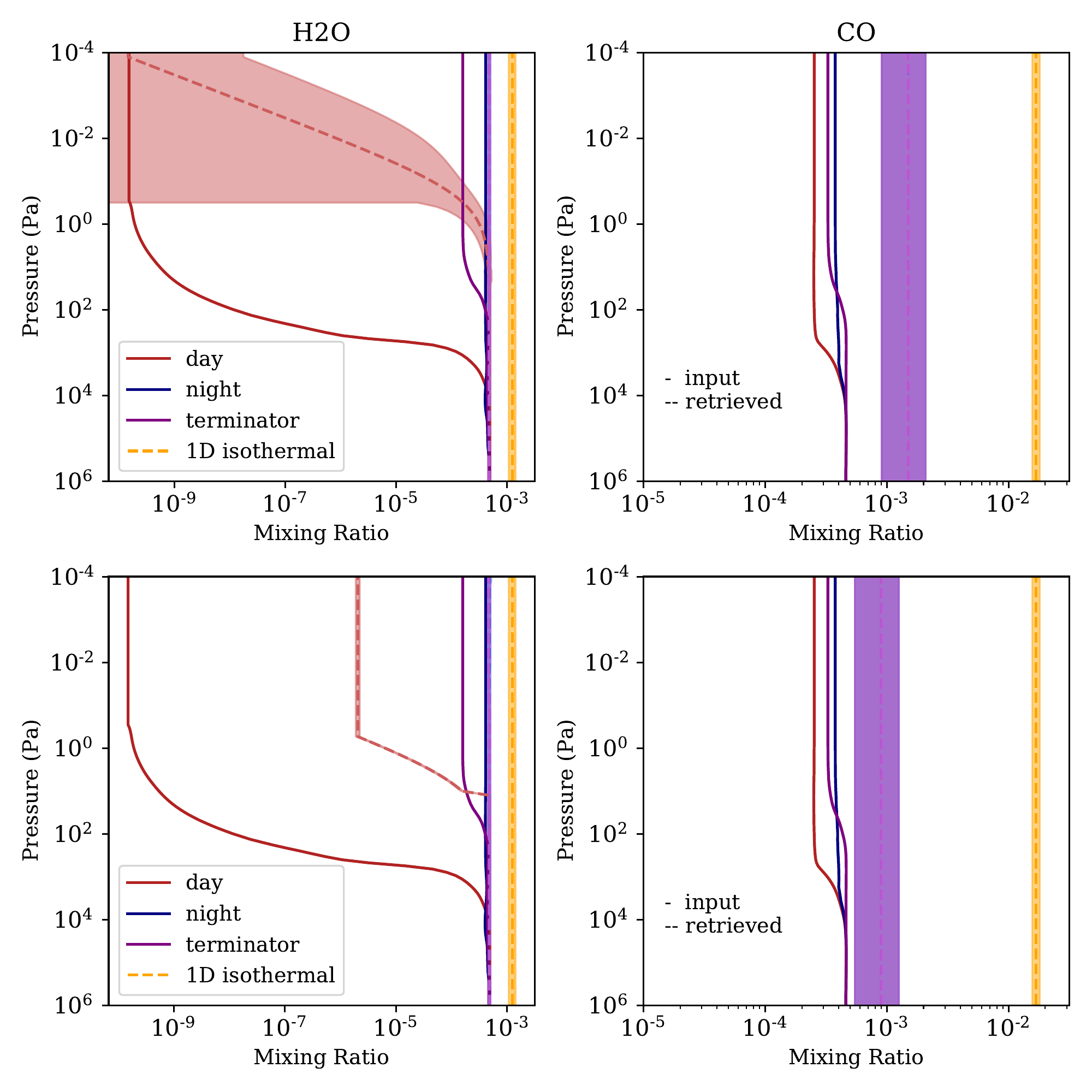}
	\caption{Mixing ratio for H$_2$O and CO for the GCM case without optical absorbers. The dashed lines are the retrieved values while the continuous lines represent the input value, In \textbf{yellow} the results from the 1D isothermal retrieval. For this retrieval, the terminator and the night mixing ratios converged to the same values.}
	\label{fig:GCMnoabs_mixing}
\end{figure}

The mixing ratio retrieved by the 2D model and the PT profile approximates the input profiles better than the 1D model as shown in figures \ref{fig:GCMnoabs_mixing} and \ref{fig:GCM_PTprofiles}. Both solutions from the two-dimensional retrieval are closer to the input mixing ratio profiles than the solution of the one-dimensional isothermal model. In particular, \taurex~2D found the correct abundance of water in the night side and the correct value of CO within 2$\sigma$ error.

To evaluate the contribution of a 2D retrieval model, we compared the solution given by a two-dimensional model with the one provided with a 1D simple model, where we also take into account molecular dissociation along the vertical dimension of the atmosphere.

From \citet{Pluriel2022}, we saw how one-dimensional models cannot interpret the input spectrum of two-dimensional models. To quantify how good the 2D model is compared to the one dimensional one, we can calculate the logarithmic Bayes factor expressed in Eq \ref{eq:logbayes_factor}.

The resulting logarithmic Bayes factor for the 2D model compared to the 1D one is $\mathscr{B} = + 89.2$, which favours the solution of the two dimensional retrieval over the one dimensional one with a $\sigma$-significance higher than $5\sigma$ \citep{trotta2008}. 

For this test case, without molecules hiding the spectral part with a significant contribution from the Rayleigh scattering, the two dimensional retrieval converges to the input spectrum compatibly with the input parameters of the synthetic spectrum. An important goal of this result is to have managed to unravel the biases observed in \citet{pluriel2020a} in the molecular abundances of the atmospheric gases. In addition, the parametrisation of our 2D retrieval model seize a thermal structure of the atmosphere which is compatible with 3D GCM models.

\subsection{Atmosphere with optical absorbers}
\begin{figure}[!htbp]
	\centering
	\noindent\includegraphics[width=\columnwidth]{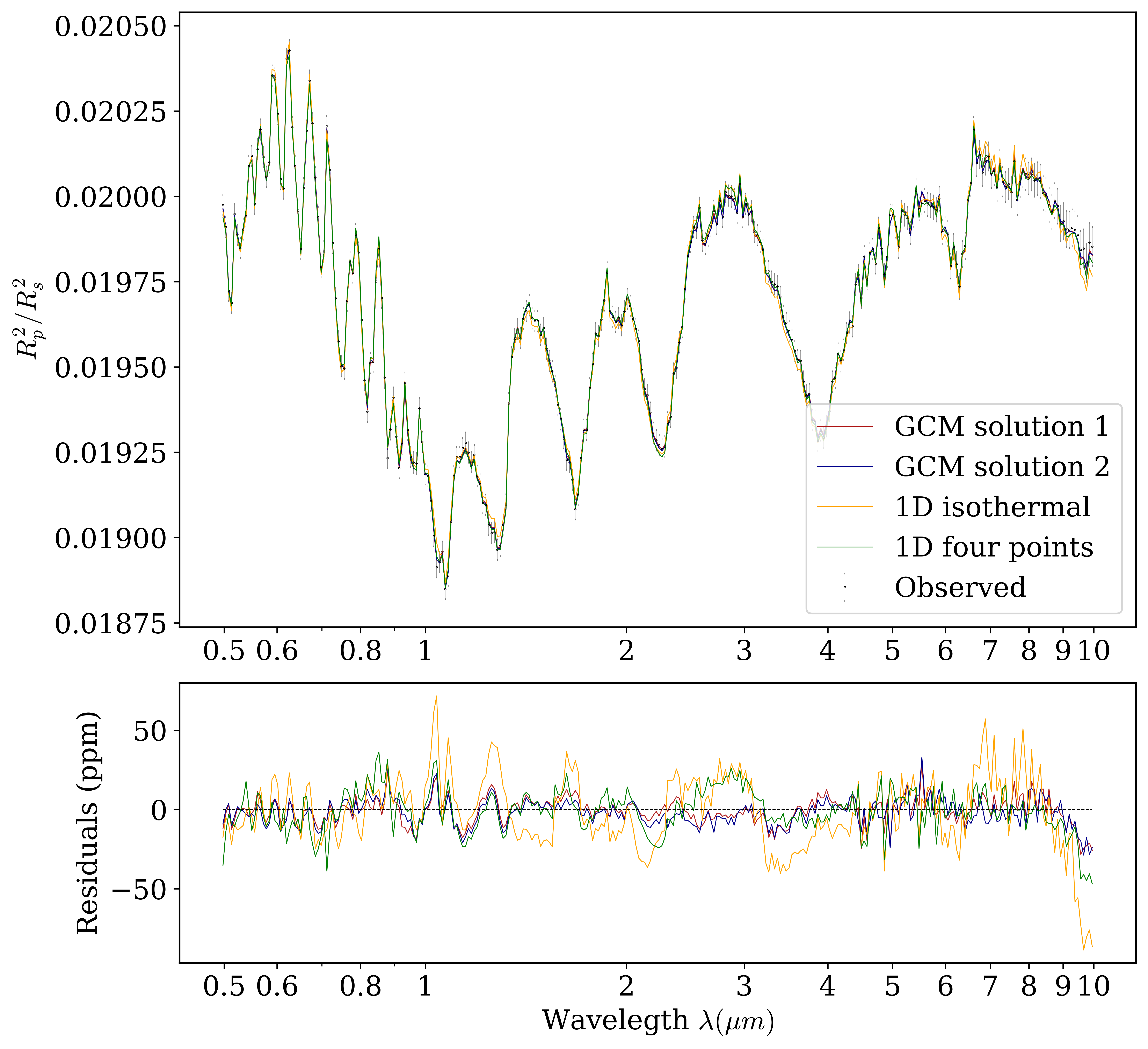}
	\caption{Best fit model for a GCM's synthetic atmosphere with TiO and VO as optical absorbers. \taurex~2D retrieval code found two possible solutions for this case (the red and the blue for the solution with, respectively, $\beta = 4.3$ and $\beta=10.0$ degrees). In \textbf{yellow} and \textbf{green} is shown the solutions from the one-dimensional retrieval using, respectively, an isothermal profile and a 4 PT point profile. \label{fig:GCM_spectrum}}
\end{figure}

After studying the case of an atmosphere without TiO and VO as optical absorbers, we want to measure how the solution of a spectral retrieval in presence of optical absorbers is impacted. Given the degeneracy of the TiO and VO and the Rayleigh scattering at short wavelengths and, depending on the spectral resolution and wavelength range of the data set, it could be common not be able to disentangle correctly the contribution of the Rayleigh scattering from the one of the optical absorbers.
The suggested Bayesian model has 11 free parameters and run in $\sim20$ days on 20 CPU cores using 1000 live points. The best fit models found by \taurex~2D is shown in Fig. \ref{fig:GCM_spectrum}.

The results are similar to the previous retrieval: (i) \taurex~2D finds two different solutions that provide the same atmospheric spectrum (ii) converging around two beta angles, $\beta_0 = 4.3_{-0.9}^{+0.9}$ and $\beta_1 = 10.0_{-0.9}^{+0.8}$, (iii) getting the expected deep abundance of each molecules considered. As we increase the degrees of freedom of the model (in our case, we have an additional spatial dimension compared to the 1D atmospheric model), the degeneracy of the solution could be a normal consequence that we need to handle. 
The $\beta_0$ angle found in this simulation is $\sim 3$ degrees higher than the $\beta_0$ solution found in the retrieval without optical absorbers. In presence of TiO and VO, we can probe a more extensive part of the day side in the high atmosphere and, then, \taurex finds a greater retrieved $\beta$ angle for this latter case.

From Fig. \ref{fig:GCM_spectrum} it is difficult to understand the contribution of the Rayleigh scattering under the TiO and VO absorption bands. In that spectral region, also the H$_2$-H$_2$ CIA provides a continuum contribution, with a peak  at around $2.2\mu$m. Campared to the case without optical absorption, we have a precision decrease from 0.02 dex to 0.10 dex, which it is related to the loss of a clear Rayleigh slope signal.

\begin{figure}[!htbp]
	\centering
	\noindent\includegraphics[width=\columnwidth]{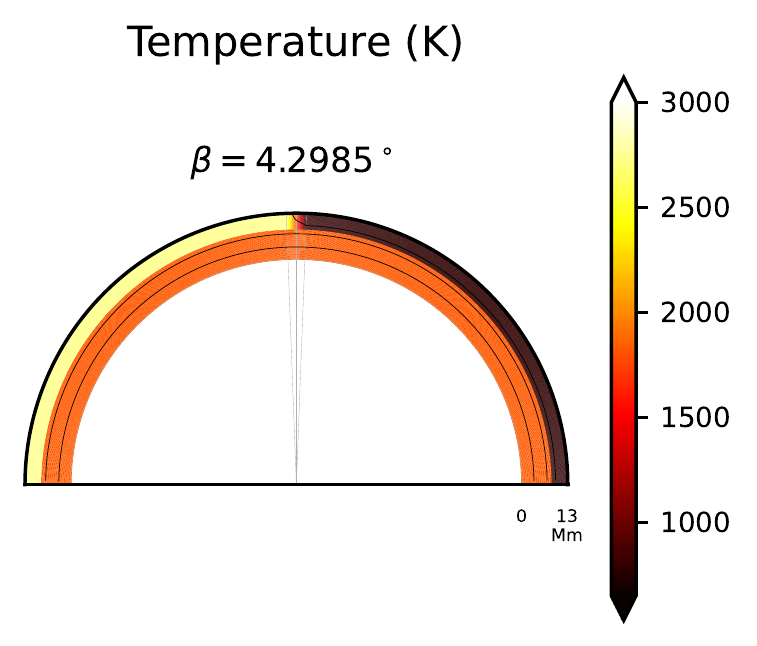}
	\noindent\includegraphics[width=\columnwidth]{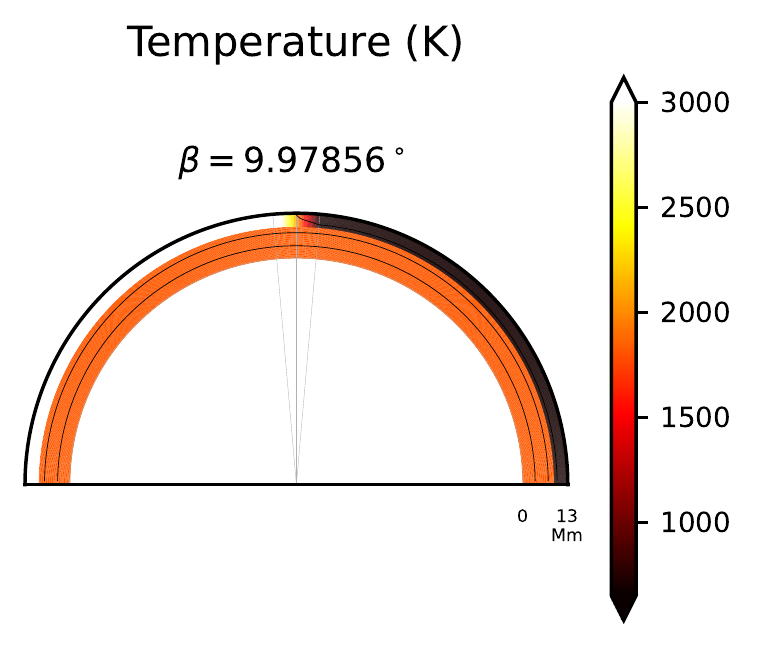}\\
	\noindent\includegraphics[width=\columnwidth]{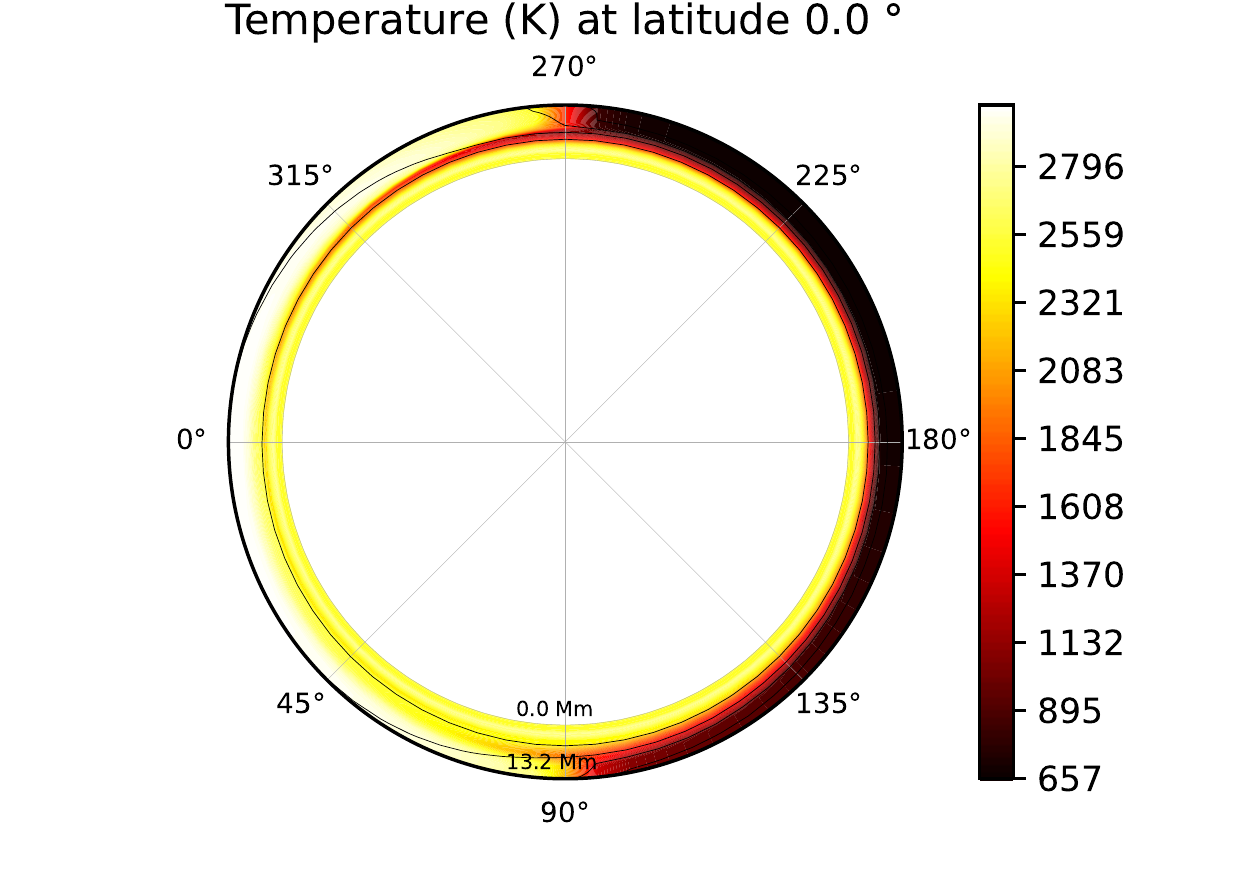}
	\caption{\textbf{Top and middle}: Temperature model for the two different solutions found in the GCM retrieval with optical absorbers. 
	For visual reasons, the size of the atmosphere has been doubled. The temperature structure on the \textbf{top} represents the one of the first solution, the green one in Fig. \ref{fig:GCMnoabs_posteriors}. On the \textbf{middle} there is the thermal structure corresponding to the second, violet solution of Fig. \ref{fig:GCM_posteriors}. \textbf{Bottom}: Temperature map (equatorial cut) of WASP-121\,b GCM simulation.
	\label{fig:GCMs_thermalstructure} }
\end{figure}

\begin{figure*}[!htbp]
    \centering
    \noindent\includegraphics[width=\textwidth]{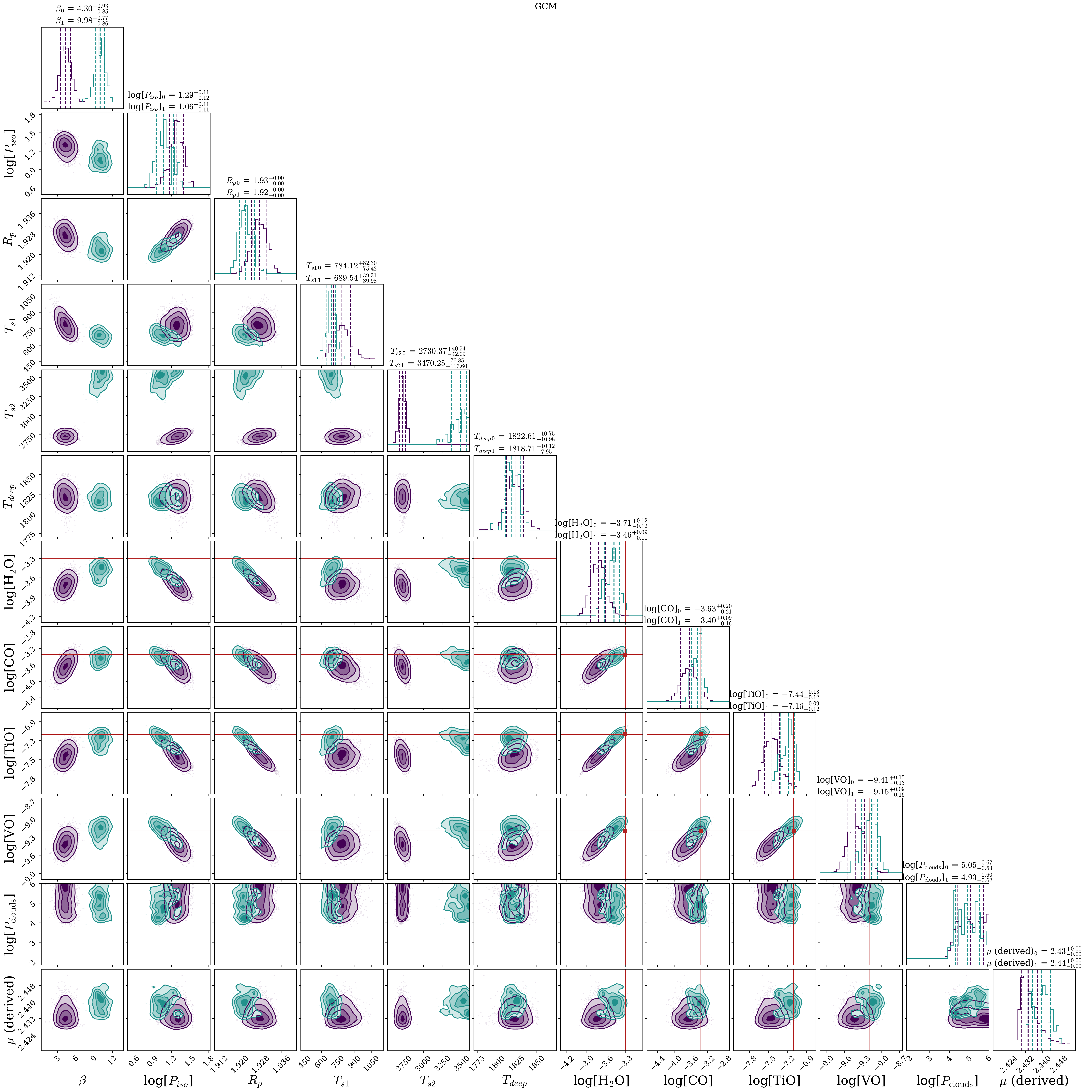}
	\caption{Best fit model for an atmosphere with TiO and VO as optical absorbers. \taurex~2D retrieval code found two possible solutions for this case. The red lines correspond to the known deep input abundances in the GCM model. \label{fig:GCM_posteriors}}
\end{figure*}

In Fig. \ref{fig:GCMs_thermalstructure} we show the retrieved thermal structure for this simulation, together with the equatorial cut view of the model. In Fig. \ref{fig:GCM_posteriors}, we can see that the two-dimensional retrieval managed to converge around the input mixing ratios for the four molecules. The temperatures retrieved are respectively $T_{night} = 780_{-80}^{+80}$K and $T_{day} = 2730_{-40}^{+400}$K for the first solution, $T_{night} = 690_{-40}^{+40}$K and $T_{day} = 3470_{-120}^{+80}$K for the second solution and $T_{deep}\sim1800$K for both solutions.

\begin{figure}[!htbp]
    \centering
    \noindent\includegraphics[width=\columnwidth]{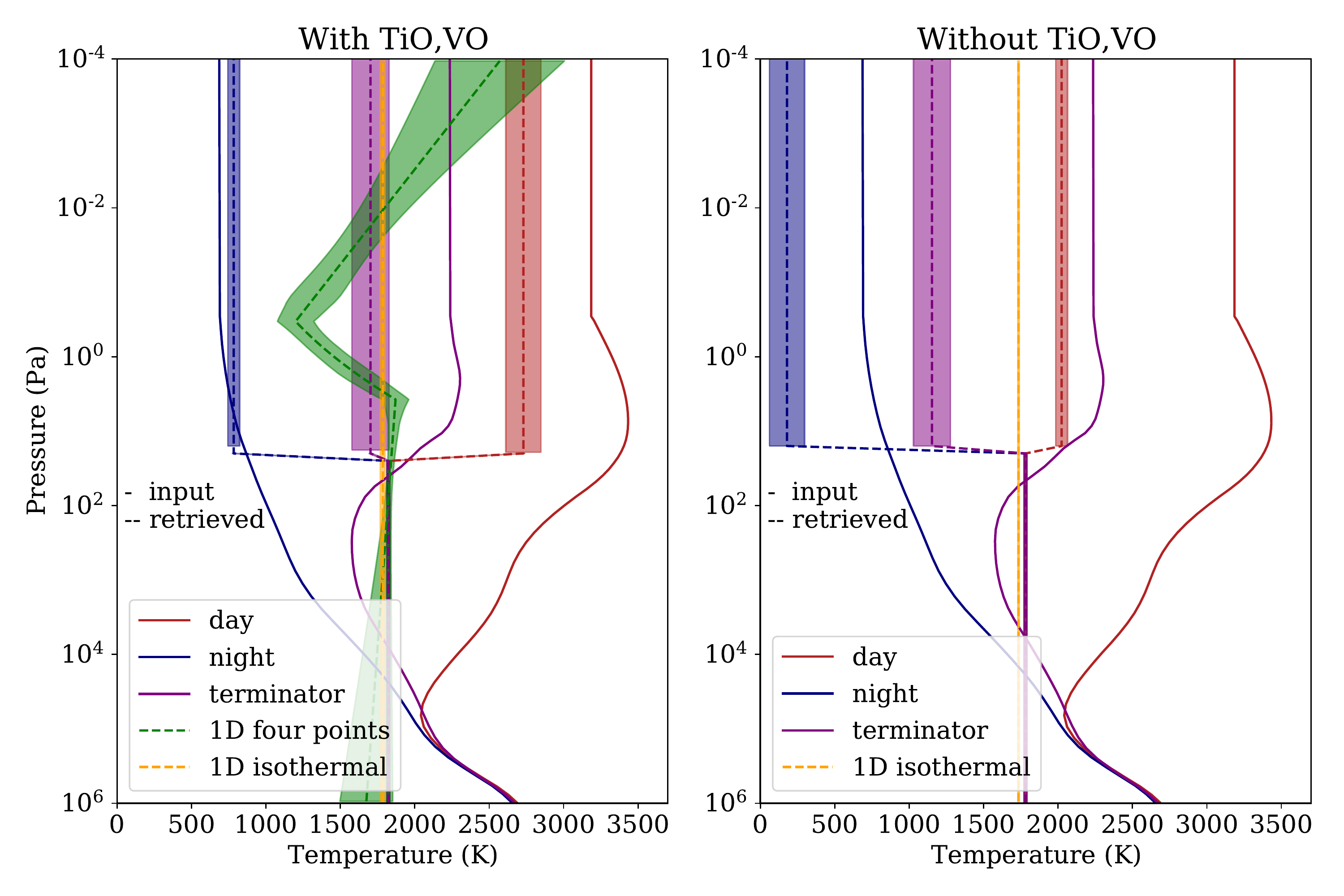}\
    \noindent\includegraphics[width=\columnwidth]{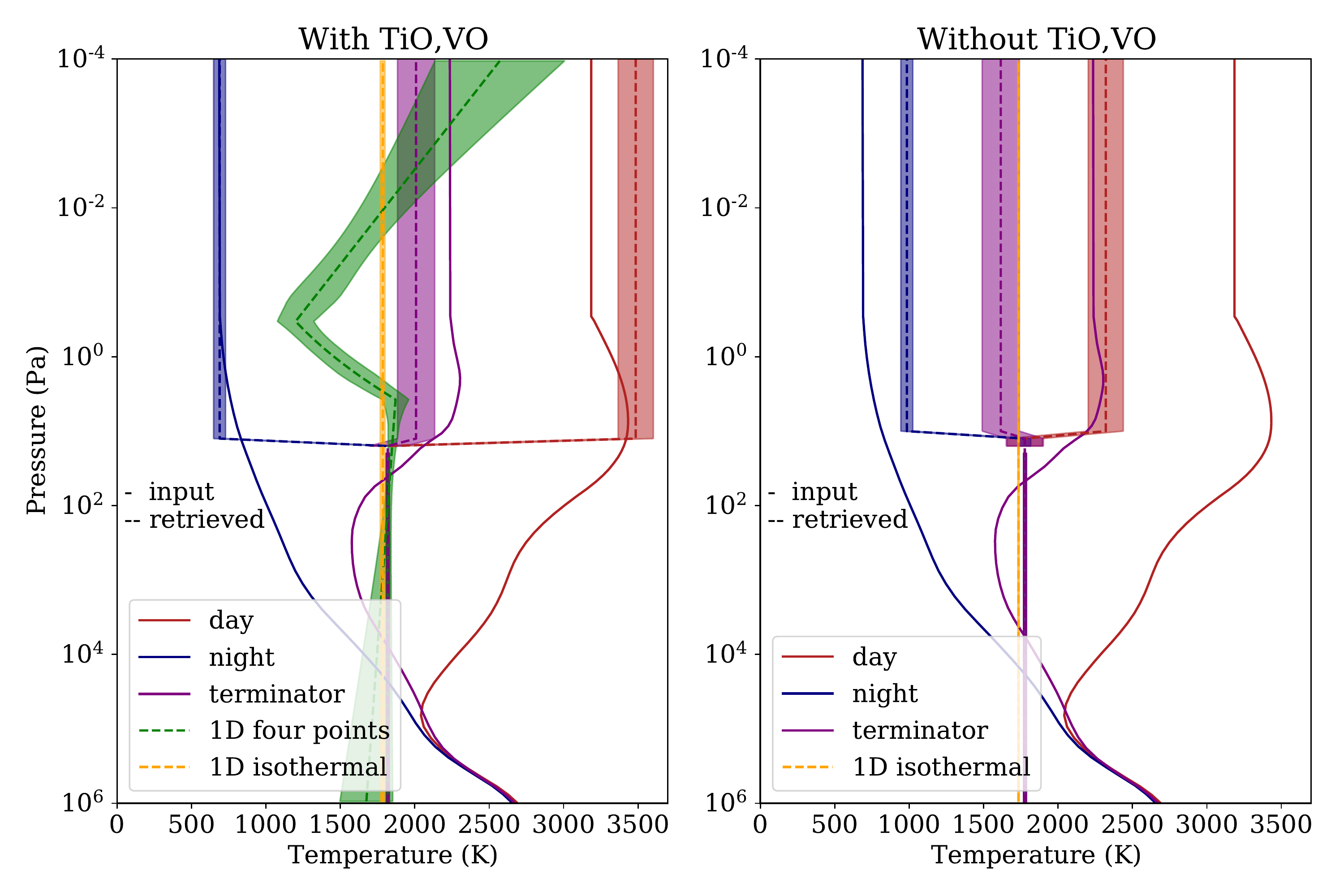}
    \caption{PT profile retrieved compared to the input PT profile. In both scenarios, with and without TiO/VO, has been used PT profile with temperature inversion. In blue, purple and red are shown, respectively, the night, terminator and day side PT profile of both the input (continuous line) and the retrieved models (dashed lines). In the upper and lower part we show, respectively, the first and the second solution from the two-dimensional retrievals. In green and yellow are shown, respectively, the solution from a one-dimensional retrieval using the 4 PT point profile (only for the GCM case with optical absorbers, and the isothermal PT profile.}. \label{fig:GCM_PTprofiles}
    
\end{figure}

\begin{figure*}[!htbp]
	\centering
	\noindent\includegraphics[width=\textwidth]{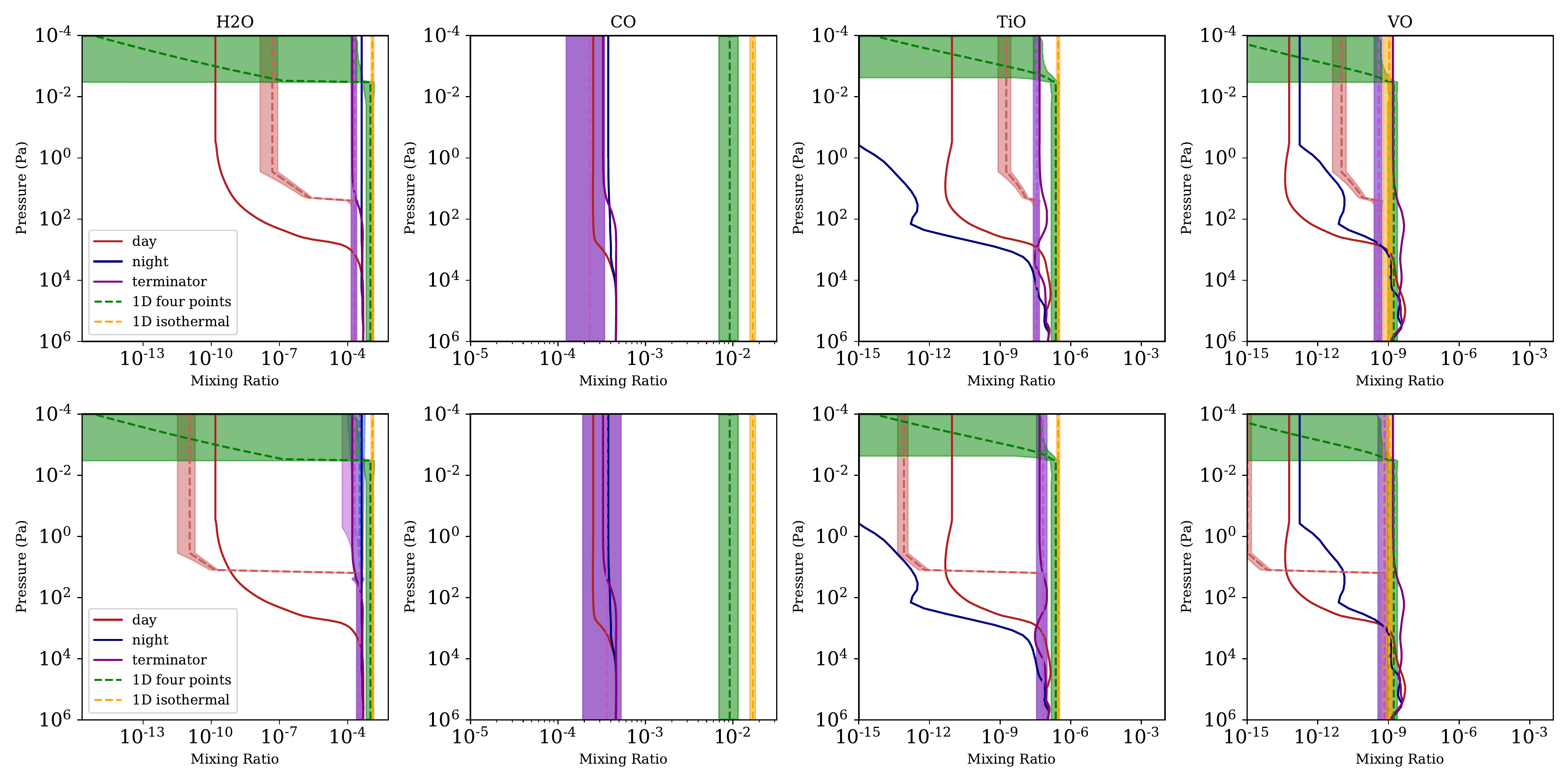}
	\caption{Mixing ratio for H$_2$O, CO, TiO and VO in the GCM test retrieval. The dashed lines are the values retrieved by \taurex while the continuous lines represent the input value. In \textbf{yellow} the results from the 1D isothermal retrieval and, in \textbf{green}, we show the solution found using a 4-point PT profile forward model. For this retrieval, the terminator and the night mixing ratios converged to the same values.}
	\label{fig:GCM_mixing}
\end{figure*}

Also in this test, the volume mixing ratio retrieved by the 2D model and the PT profile approximates better the input profiles than the 1D retrieval, as shown in figures \ref{fig:GCM_PTprofiles} and \ref{fig:GCM_mixing}. \taurex~2D found two solutions, approximating the input day and night temperatures and estimating the deep atmosphere around its average temperature (Fig. \ref{fig:GCM_PTprofiles}). Moreover, \taurex~2D found a chemical distribution of elements closer to the input distribution than the one found with the one dimensional retrievals (Fig. \ref{fig:GCM_mixing}). 


The crucial result of this test is that \taurex~2D manage to retrieve the expected deep abundance of each molecule with a very good agreement to fit the input data. \taurex~2D correctly separate the contribution of the Rayleigh scattering from the one of the optical absorbers, unraveling also here the biases explained in \citet{pluriel2020a}.

Also in this case, a comparison with an equivalent 1D retrieval leads to a Bayes factor $\mathscr{B} = + 47.9$ in favor of the 2D retrieval.

\section{Comparison with a more complex 1D model retrieval}
\label{comparison1D}

The 2D retrieval, in this case, helps us digging more on the three-dimensional structure of the planet. We see that the two different solutions differ mainly from two possible $\beta$ angles which lead to two possible values of $T_{day}$. We remark that a higher $\beta$ angle results in a higher day-side temperature.

To quantify the goodness of the fit, and see the advantages of a two-dimensional retrieval over the "classic" one-dimensional, we fitted the same input spectrum with a one dimensional model and calculated the Bayes factor between the two best-fit models.

We tested the one-dimensional model twice, assuming two different pressure-temperature profiles: one isothermal atmosphere and \textbf{one} four-point PT profile atmosphere.

\begin{table*}[!htbp]
	\centering
	\caption{Comparison of models\label{tab:1Dcomp}}
	\begin{tabular}{|c|c|c|c|c|c|}
		\hline
		 Retrieval type & Number of parameters & Log Bayesian Evidence & $\tilde{\chi}^2$ & $\mathscr{B}$ with 1D iso & $\mathscr{B}$ with 1D 4-point PT \\
		\hline
		2D sol$_0$ & 11 & 2795.0 & 0.07 & 47.9 & 14.5 \\
		\hline
		2D sol$_1$ & 11 & 2795.0 & 0.07 & 47.9  & 14.5\\
		\hline
		1D 4 PT & 12 & 2780.5 & 0.18 & 33.4 & -\\
		\hline
		1D iso & 7 & 2747.1 & 0.45 & - & -33.4\\
		\hline
	\end{tabular}
\end{table*}

In Tab \ref{tab:1Dcomp} we can find a resume of all the comparison made between the 2D retrieval including TiO and VO and the 1D retrievals. Together with the Bayesian evidence, we also provide the reduced $\tilde{\chi}^2$ of each best model which is defined as follows:

\begin{equation}
\tilde{\chi}^2 = \frac{1}{N - p} \sum_{i=1}^N \left( \frac{O_i - C_i}{\sigma_i} \right)^2,
\end{equation}

Where $O$ and $C$ are, respectively the observed and the computed spectral points, $\sigma$ is the uncertainty, $p$ is the number of parameters in the model and $N$ is the total number of measurements. As described in Tab \ref{tab:1Dcomp}, the 2D models approximate better the input spectrum, returning the lowest $\tilde{\chi}^2$ with respect with the 1D retrievals.

In Fig. \ref{fig:GCM_spectrum} we can see that the four PT points retrieval leads to a better retrieval than the one done using an isothermal atmosphere. The logarithmic Bayes factor $\mathscr{B}= \log{E_{4pt}} - \log{E_{iso}} = 33$, which confirms that the four-points retrieval is statistically better than the isothermal retrieval with a $\sigma$-significance higher than $5\sigma$.

A comparison between the two dimensional and the four-points retrieval leads to a logarithmic Bayes factor of $\mathscr{B}= \log{E_{2D}} - \log{E_{4pt}} = 14$ and, then the two-dimensional retrieval is better also than the four points retrieval with a $\sigma$-significance higher than $5\sigma$.

The reduced $\tilde{\chi}^2$ of the four points retrieval tells us that the atmospheric model found is compatible with the input spectrum. 

All the molecular abundances appear to be overestimated, and then, biased compared to the input parameters used to generate the input spectrum.

\begin{figure*}[!htbp]
    \centering
    \noindent\includegraphics[width=\textwidth]{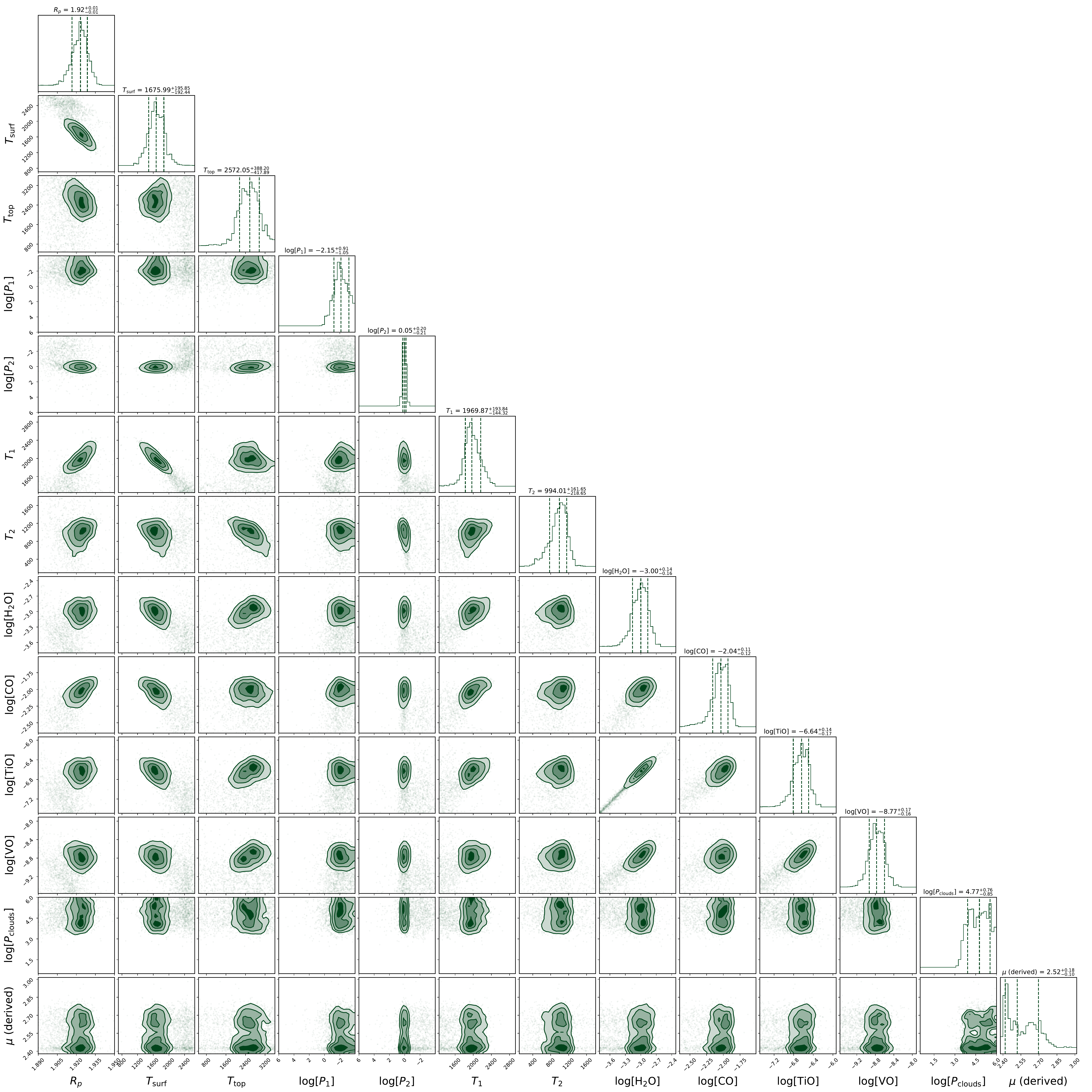}
    \caption{Posterior distributions related to the four-points retrieval. Confirming the parameter biases found by \citet{Pluriel2022} \label{fig:oned_posteriors}.}
\end{figure*}

The calculation of the Bayes factor allows us to quantify the goodness of the 2D retrieval compared to the one-dimensional one. Indeed, if we consider only the reduced $\tilde{\chi}^2$ of the fit, we see that the single models are compatible with the input spectra with a high significance. However, it is impossible to get rid of the parameters biases when using the one-dimensional model. The key-point of the calculation of the logarithmic Bayes factor is to highlight the fundamental differences of the two model and the limits given by their geometry. With 30ppm error bars it is, indeed, possible to reveal and explain some spectral features by introducing an additional geometric dimension on the atmospheric model.

Indeed, even though the one-dimensional retrieval converges toward a statistically significant atmospheric model, the parameters found can differ from the ground truth by several orders of magnitudes. 
In Fig. \ref{fig:oned_posteriors}, as an example, we show the posterior distributions for a retrieval of the GCM input spectrum with optical absorbers, using a one-dimensional atmospheric model and a four temperature-pressure points. The best model fits very well the input spectrum, but the retrieved amount of atmospheric trace gases differ from the ground truth up to 2 orders of magnitudes.

We see that, the logarithmic Bayes factor defined in Eq \ref{eq:logbayes_factor}, that expresses the difference of the 2D model compared with the 1D one is $\mathscr{B} = + 47.9$, which privileges the solution of the bi-dimensional model with a $\sigma$-significance higher than $5\sigma$. Also in this case the 2D model leads to a statistically better fit than the one-dimensional model.

The statistical significance of the 2D model is higher than that of the 1D model, and it is less affected by the parameters biases, since it approximates better the three-dimensional geometry of the input model.

The comparison with the one-dimensional retrieval emphasizes the importance of running more-than-one dimensions retrieval models when using 30ppm noise input spectra. 

In the era of the JWST, ARIEL and the Twinkle space missions it will be fundamental to use better models which rely on more complex geometrical assumptions than the plane parallel atmospheres.

\section{Conclusion}

More-than-one dimensional retrievals will be crucial to better interpret the input spectral observation of new generation instruments. 
Even though the approximation of a plane-parallel atmosphere can lead to a statistically acceptable fit, it inevitably leads to parameters biases that can be very far away from the reality \citep{pluriel2020a,Pluriel2022}.
This is especially true for HJs and UHJs, where the great dichotomy between the day and the night side of the planet may lead to a misinterpretation of the input observations, without an accurate model.

We developed a 2D version of \taurex, which demonstrates that a two dimensional forward model is a good compromise between model accuracy and computational requirements. This model can solve the issues of the parameters biases induced by a one dimensional model which is essential in the context of future mission such like JWST, Twinkle or Ariel. Thanks to \taurex~2D we are able to get with much higher confidence than with 1D retrieval models the abundances of the species.

We showed how the model choice could be crucial to properly retrieve the thermal structure of an exoplanetary atmosphere (in Section \ref{symmetry}), due to the intrinsic degeneracies of a high dimensional transmission model.

Finally, in Section \ref{comparison1D} we demonstrated how a two-dimensional retrieval with an isothermal atmosphere returns a statistically more significant solution than a one-dimensional retrieval using a more complex 4-points pressure temperature profile. 

\begin{acknowledgements}
This project has received funding from the European Research Council (ERC) under the European Union's Horizon 2020 research and innovation programme (grant agreement n$^\circ$679030/WHIPLASH). TZi acknowledges the funding support from Italian Space Agency (ASI) regulated by `Accordo ASI-INAF n. 2013-016-R.0 del 9 luglio 2013 e integrazione del 9 luglio 2015 CHEOPS Fasi A/B/C'.
\end{acknowledgements}

\bibliographystyle{aa}
\bibliography{biblio}

\end{document}